\documentclass[11pt]{article} 

\usepackage[utf8]{inputenc} 

\usepackage{graphicx} 

\usepackage{amsfonts}

\usepackage{mathtools}
\usepackage{booktabs} 
\usepackage{array} 
\usepackage{verbatim} 
\usepackage{subfig}
\usepackage{amsmath}
\usepackage{natbib}
\usepackage{epstopdf}
\usepackage{tikz}
\usetikzlibrary{arrows,decorations.markings}

\usepackage[margin = 1in]{geometry}
\title{Modelling the spatial extent and severity of extreme European windstorms}

\author{Paul Sharkey$^{1}$, Jonathan A. Tawn$^{2}$ and Simon J. Brown$^{3}$}

\date{} 

\begin{document}
\maketitle
\vspace{-40pt}
\begin{center}
Email: \texttt{pgshky@gmail.com}, \texttt{j.tawn@lancs.ac.uk}, \texttt{simon.brown@metoffice.gov.uk} \\
$^{1}$\emph{British Broadcasting Corporation, Bridge House, Salford Quays, Salford, M50 2LH, U.K.} \\
$^{2}$\emph{Department of Mathematics and Statistics, Lancaster University, Lancaster, LA1 4YF, U.K.} \\
$^{3}$\emph{Met Office Hadley Centre, Exeter, EX1 3PB, U.K.}
\end{center}

\begin{abstract}
Windstorms are a primary natural hazard affecting Europe that are commonly linked to substantial property and infrastructural damage and are responsible for the largest spatially aggregated financial losses. Such extreme winds are typically generated by extratropical cyclone systems originating in the North Atlantic and passing over Europe. Previous statistical studies tend to model extreme winds at a given set of sites, corresponding to inference in a Eulerian framework.
Such inference cannot incorporate knowledge of the life cycle and progression of extratropical cyclones across the region and is forced to make restrictive assumptions about the extremal dependence structure. We take an entirely different approach which overcomes these limitations by working in a Lagrangian framework.
Specifically, we model the development of windstorms over time, preserving the physical characteristics linking the windstorm and the cyclone track, the path of local vorticity maxima, and make a key finding that the spatial extent of extratropical windstorms becomes more localised as its magnitude increases irrespective of the location of the storm track. Our model allows simulation of synthetic windstorm events to derive the joint distributional features over any set of sites giving physically consistent extrapolations to rarer events.  From such simulations improved estimates of this hazard can be achieved both in terms of intensity and area affected.    
\end{abstract}
\hspace{3.5mm}{\bf Keywords:} Climate extremes, extratropical cyclones, extreme value analysis, Lagrangian model, spatial dependence.

\section{Introduction}
\label{sec:intro}
While the winter climate of the United Kingdom and northern Europe is typically associated with mild, wet weather that poses little infrastructual or societal risk, there has been an increased focus in recent years on the impact of windstorms in this part of the world. These events are often the consequence of extratropical cyclones, and are directly linked to the occurrence of flooding, transport chaos and considerable damage to infrastructure. \citet{roberts2014xws} describe a comprehensive catalogue of European windstorms in the period 1979-2012 that contains extensive information related to the meteorology and monetary impact of each storm. Storm Daria, which occurred in January 1990, is believed to be the most destructive windstorm in this period, with an estimated insured loss of $\$8.2$bn. \\

Windstorms are often a consequence of the passage of extratropical cyclones. Extratropical cyclones are synoptic-scale weather systems associated with low pressure that generally originate in the North Atlantic and progress northeasterly towards Europe. These systems can be characterised by paths of local vorticity maxima, which we refer to as tracks. Cyclones are typically formed as a result of horizontal temperature gradients and evolve according to a particular lifecycle with  associated frontal systems \citep{shapiro1990fronts} where cold and warm air masses converge. High winds tend to occur along these boundaries \citep{hewson2015cyclones}. A large body of research exists on cyclone identification, storm tracking and feature extraction in reanalysis datasets \citep{murray1991numerical,hodges1995feature}, which produce good approximations of how a track develops in space and time. The study of cyclones with respect to the cyclone centre, a Lagrangian frame of reference, has provided useful insight into the evolution of cyclones and the physical process that drive evolution \citep{catto2010, rudeva2011, dacre2012}.  However, there has been little work on characterising the ensuing extreme winds in a robust way within such a framework. \\

We would like to assess the joint risk of multiple locations, in a set of fixed sites, experiencing the same windstorm event.  This assessment is particularly difficult when using only the data from these sites as this cannot  exploit knowledge  that these extreme events are arising from the passage of extratropical cyclones over time and space. Consequently, inference cannot account for extreme events that, just by chance, did not pass over the observed sites and it does not incorporate the knowledge of the properties of extratropical cyclones into the estimation of probabilities of rarer events than those observed. Our paper describes an approach which incorporates this physical knowledge through a statistical model of windstorms, which can be used in practice to assess the marginal and joint risk of these weather systems over Europe while accounting for the varying probability of storm tracks over the region.\\

A common approach to statistical modelling of extreme wind speeds is to use techniques from extreme value analysis, which use models built on asymptotic arguments to estimate probabilities of events beyond the range of the data. In meteorological applications, such probabilities are commonly used by practitioners to design infrastructure appropriately to defend against the natural hazard being studied. The most widely-used approach in extreme value analysis is to consider excesses above a suitably high threshold. Consider a sequence of independent and identically distributed (i.i.d.) random variables $X_1, \hdots, X_n$. Under weak conditions on the $X_i$, the unique, non-degenerate distribution that the scaled excesses of a threshold by $X_i$ converges to, as the threshold tends to the upper limit $x_F$ of $X_i$, is the generalised Pareto distribution (GPD) \citep{pickands1975statistical,davison1990models}. We make the assumption that this limiting result holds for a large enough threshold $u$. The GPD takes the form
\begin{equation}
\Pr(X_i-u > x | X_i > u) = {\left(1+\frac{\xi x}{\sigma_{u} }\right)}^{-1/\xi}_{+}, \mbox{                                } x > 0
\label{eq:gpd}
\end{equation}  
where $c_{+} = \max(c,0)$ and where $\sigma_u > 0$ and $\xi \in \mathbb{R}$ denote the scale and shape parameters respectively. The shape parameter is invariant to the choice of threshold but the scale parameter is threshold-dependent. The threshold $u$ is typically determined using some standard selection diagnostics \citep{coles2001introduction} such as ensuring that the parameters are stable with respect to the threshold for all threshold choices larger than $u$. \\

There have been numerous studies using extreme value models to estimate extreme wind speeds \citep{coles1994directional,fawcett2006hierarchical,ribatet2013spatial}. However, these models have no consideration of the physical processes generating the extremes. Some recent studies have, however, modelled extreme winds in the context of an extratropical cyclone. \citet{della2009statistical} use a GPD model to assess changes in extreme wind intensity under climate change scenarios, which led to results showing that that the frequency of intense wind events in Europe is predicted to increase. \citet{sienz2010extreme} extended this approach to model the effect of the North Atlantic Oscillation (NAO) index. \citet{bonazzi2012spatial} modelled the tail dependence of wind speeds between locations over Europe using a bivariate extreme value copula and found dependence to be greater in the west-east direction, which is consistent with the passage of extratropical cyclone tracks over Europe. More recently, \citet{youngman2016geostatistical} use extreme value analysis coupled with a geostatistical model to capture the spatio-temporal development of windstorms over Europe, but again the direct influence of the storm track is not accounted for. \\

The existing approaches share a common philosophy in that windstorms are modelled in an Eulerian frame of reference. In fluid mechanics, this approach refers to the scenario whereby an observer measures observations of a process at a fixed location while the process, e.g., a windstorm, passes over. So, it is a Eulerian framework that is being adopted when analysing data collected at a fixed set of sites over time, such as when modelling weather station data. It is the standard approach for statistical modelling of spatial data \citep{cressie1993statistics}, with the advantage that one can build large data sets, with time series at each location being observed. However, if one's concern is focused more on modelling the evolution and influence of the process itself, a Lagrangian frame of reference is required. This approach to modelling requires following the process and collecting observations as it moves through space and time. We take this approach as it is a natural framework on which to build a model for windstorms, since it allows us to model the behaviour of extreme winds relative to the storm centre. \\

Recent advances in climate modelling have resulted in the increased availability of high-resolution datasets that are spatially and temporally complete, allowing large-scale processes to be modelled in a Lagrangian framework \citep{catto2010}.  As the observed data record is relatively short with regard to storm tracks, and even more so with regard to windstorms, there is a need for a statistical model to provide extra information about the possible extreme, long-term characteristics of windstorms that are generated by the extratropical cyclone. In particular, we need to be able to assess the likelihood of observing more severe storms than those observed, where these might occur, and how large the spatial extent of the event might be. \\

\citet{sharkey2018thesis} and \citet{sharkey2017tracking} model synthetic storm tracks in a way that replicates the climatology of extratropical cyclones in the North Atlantic. This model allows extrapolation to events that have larger vorticity than previously observed. We extend this work to model synthetic windstorm events relative to their storm tracks in a Lagrangian framework. We first describe a model for the instantaneous area, which we refer to as a footprint, affected by strong winds in the vicinity of the storm centre. We represent the footprint as an ellipse 
as this provides a parsimonious envelope for typical event shapes. We then present an algorithm which determines the location and shape of the footprint at any point in time. We model the evolution over time of both the characteristics of the ellipse and the magnitude and spatial distribution of the extreme winds within the footprint. These models allow us to generate a series of footprints for multiple windstorms associated with the synthetic storm tracks of \citet{sharkey2017tracking}, providing a method for estimating the entire area affected by extreme winds from the passage of a single storm and the aggregated risk arising from extreme windstorms over the North Atlantic and Europe. \\

The paper is structured as follows. In Section~\ref{sec:data}, we introduce the data and our methods for extracting the features of the windstorm from the data and an exploratory analysis of these features. We discuss our modelling strategy in Section~\ref{sec:methodology}, introducing our approaches to modelling the evolution of the windstorm footprints and the extreme winds within the footprints. We derive a range of estimated spatial properties of windstorms over Europe in Section~\ref{sec:results}, before concluding in Section~\ref{sec:discussion} with some discussion.

\section{Windstorm definition and exploratory analysis}
\label{sec:data}

\subsection{Data description}
\label{subsec:desc}
As in \citet{sharkey2017tracking}, our work uses storm track data covering the North Atlantic and European domain. Our dataset consists of storm track locations at 3-hourly intervals with an associated vorticity measure representing the strength of the storm. Storms are identified and tracked over the period 1979-2014 from the ERA-Interim reanalysis dataset \citep{dee2011era} using a feature extraction approach outlined in \citet{hoskins2002new} based on the tracking algorithm introduced in \citet{hodges1995feature}. We restrict our attention to the set of storm tracks produced during an extended winter period (October-March), when storms are widely regarded to be most intense. We exclude Mediterranean extratropical cyclones as these do not produce large financial losses. We also exclude ``medicanes" \citep{akhtar2014medicanes} as these arise from a different physical processes and are not captured well by reanalysis data. We denote the longitude and latitude coordinates of the storm track at time $t$ by $\text{Lon}_t$ and $\text{Lat}_t$ respectively. The vorticity associated with the track at $(\text{Lon}_t, \text{Lat}_t)$ is denoted by $\Omega_t$. \\

Our model is based on wind speed data from the EURO4 numerical weather prediction model \citep{standen2017prediction}, which is a downscaled version of the ERA-Interim reanalysis dataset. Data are available on a 4 km spatial resolution over Europe and part of the North Atlantic, amounting to $1,100,000$ cells (see Figure~\ref{fig:gpd_par}). Values are obtained at hourly intervals over the period 1979-2014. We linearly interpolate the storm track locations and vorticity within each 3-hourly interval to match the hourly temporal resolution of the wind speed data. We select only the wind speed fields at times corresponding to the set of storm tracks. In particular, as we are looking to model the effect of the storm track on the spatio-temporal evolution of wind speeds in the vicinity of the track, we isolate the field of interest as a square-shaped region centred at the storm centre with sides of approximately $1,600$ km in length (see Figure~\ref{fig:marg_std}, which in the left panel shows such a region at a time step when storm Daria was located over the UK). We believe that this field is large enough so that the extreme winds generated by a windstorm are sufficiently captured. 

\subsection{Marginal model}
\label{subsec:wind_margins}
\begin{figure}[h!]
\centering
\includegraphics[width=15cm]{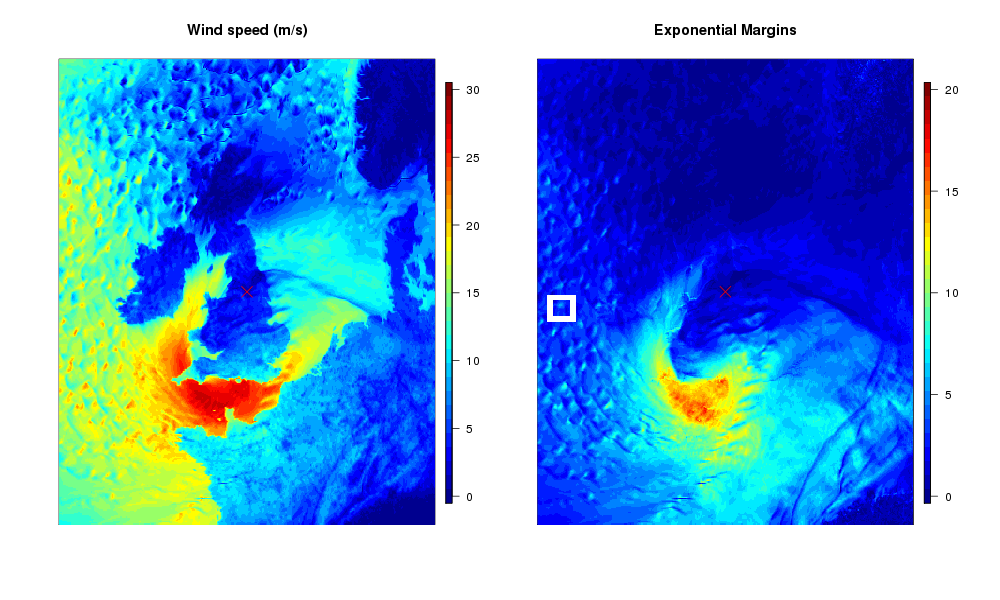}
\caption{Wind speeds, in m/s, at $3$pm on January 25th, 1990 in the vicinity of Storm Daria (left) and standardised onto Exp$(1)$ margins (right). The storm centre is represented by the cross. The white box contains an example of a localised convective event. Land/sea borders are not explicitly shown in the left panel, but can be seen due to the contrast in magnitude between winds over land and sea. }
\label{fig:marg_std}
\end{figure}

Initial investigation of the data confirms, as expected, that winds over the sea are markedly stronger than those over land (see Figure~\ref{fig:marg_std}, left panel). This property is largely due to open water exerting significantly less drag on the atmosphere in contrast with the land surface, orography and man-made structures that impede strong winds. The contrast in scale over land and sea, and to a lesser extent, over low-lying and high-lying land, motivates a standardisation of wind speeds in each cell to have a common marginal distribution. In this sense, our approach is based on a copula modelling strategy over space \citep{joe1997multivariate}. \\

Let $X(s,t)$ be a random variable denoting the wind speed in cell $s$ at time $t$, for $s=1,\hdots,1,100,000$. We propose a marginal model for the distribution of $X(s,t)$, for each cell $s$, of the form
\begin{equation} F_s(x) = \begin{dcases*} 
							\hat{F}_s (x) & $x \leq u_s$ \\
 1 - \lambda_{u_s} {\left( 1 + \xi_s \frac{x - u_{s}}{\sigma_{u_s}} \right)}^{-1/\xi_s}  &  $x > u_s,$
										 \end{dcases*} 
\label{eq:marginal_model}
\end{equation}
where $\hat{F}_s$ denotes the empirical distribution function for realisations of $X(s,t)$ over time. For realisations above $u_s$, the GPD in~(\ref{eq:gpd}) is used as a conditional model for excesses above $u_s$, with cell-specific parameters $(\sigma_{u_s},\xi_s)$. To undo this conditioning, a third parameter $\lambda_{u_s}$, denoting the probability of an exceedance of $u_s$, must be specified. Parameter stability plots were checked at a number of cells, which indicated that a threshold corresponding to the $98\%$ marginal quantile in each cell would be a good choice for all cells. We therefore choose this quantile to be the cell-specific threshold in each cell. Parameter estimates are obtained using maximum likelihood techniques. We note that we do not attempt to impose spatial smoothness on the form of $F_s$. \\
\begin{figure}
\centering
\includegraphics[width=17cm]{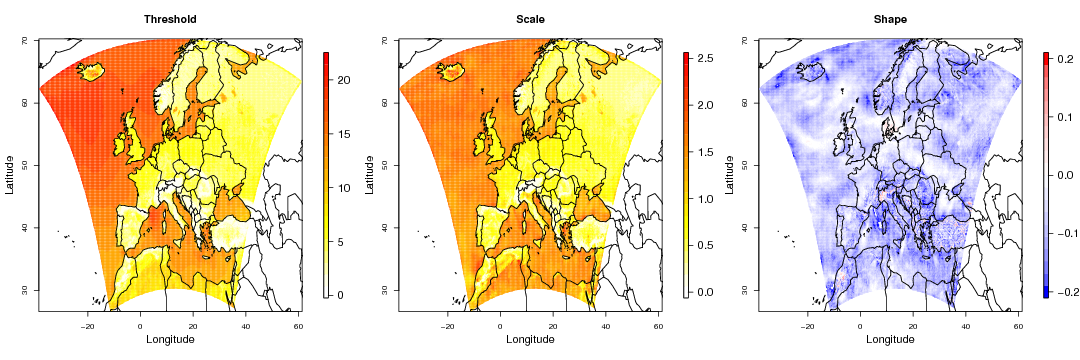}
\caption{The extreme value threshold (left) used in the specification of the GPD model, along with parameter estimates of the scale (centre) and shape parameters (right) for each cell.}
\label{fig:gpd_par}
\end{figure}

Figure~\ref{fig:gpd_par} shows the parameter estimates of the GPD corresponding to each cell in the region over Europe along with the threshold, which corresponds to the $98\%$ quantile in each cell. This figure shows explicitly the contrast in wind speed magnitudes between locations on land and sea. This contrast is also reflected in the estimation of the scale parameter, but the shape parameter exhibits no such contrast between land and sea, with most estimates occurring in the region $(-0.2,0)$, indicating that the distribution of wind speeds has a finite endpoint in general. The numerical maximisation algorithm used to obtain the parameter estimates mostly converges, however there are certain regions that exhibit unusual behaviour. For example, the $98\%$ quantile threshold is a lot higher in areas over Iceland than other land locations, while the Italian Alps see unusually high estimates of the shape parameter. The challenges of representing wind over high orography in numerical weather models is well documented, as evidenced by these two particular locations, and recalibration methods have been proposed \citep{howard2007correction}. For this analysis we, exclude regions with orography above $500$m from our analysis. \\

We propose a marginal standardisation to unit exponential Exp$(1)$ margins using a probability integral transform using the marginal model~(\ref{eq:marginal_model}). The exponential scale has been found to be most ideal for studying extremal dependence \citep{heffernan2004conditional,wadsworth2013new}. We define $X^{E}(s,t)$ to be the relative wind speed on Exp$(1)$ margins in cell $s$ and time $t$, such that for all cells and all times we have that
\[ X^{E}(s,t) = -\log\{1-F_s (X(s,t))\}, \]
where $F_s$ is defined as in~(\ref{eq:marginal_model}).
We use the term relative wind speeds to describe $X^{E}(s,t)$ as this quantity defines wind speeds relative to the marginal characteristics of cell $s$. 
Figure~\ref{fig:marg_std} shows the effect of the standardisation on one time step of Daria. In particular, we see spatially correlated values of high relative wind speeds over both land and sea as a result of the transform and the land/sea contrast is almost entirely removed. Thus, this transformation recovers the underlying process of the windstorm. This approach reveals the shape of the windstorm event without the influence of the marginal characteristics at each location, which should allow for a simpler approach to modelling the spatial extent of the event.

\subsection{Feature extraction}
\label{subsec:feature}
Figure~\ref{fig:marg_std} shows that, as well as the large band of strong relative winds clustered near the storm centre, small localised fragments of high relative wind speeds are visible on the western edge of the footprint. Such localised events are due to convection and not due to the larger scale dynamics of the storm. Since we believe them not to be directly linked with the extratropical storm system, we are not concerned with modelling these features. Our work is focused on modelling the features of an extratropical cyclone that occur on larger spatial scales and have the potential to produce much larger impacts, such as large insurance losses \citep{roberts2014xws}, than these localised convective events. \\

\begin{figure}[h!]
\centering
\includegraphics[width=16cm]{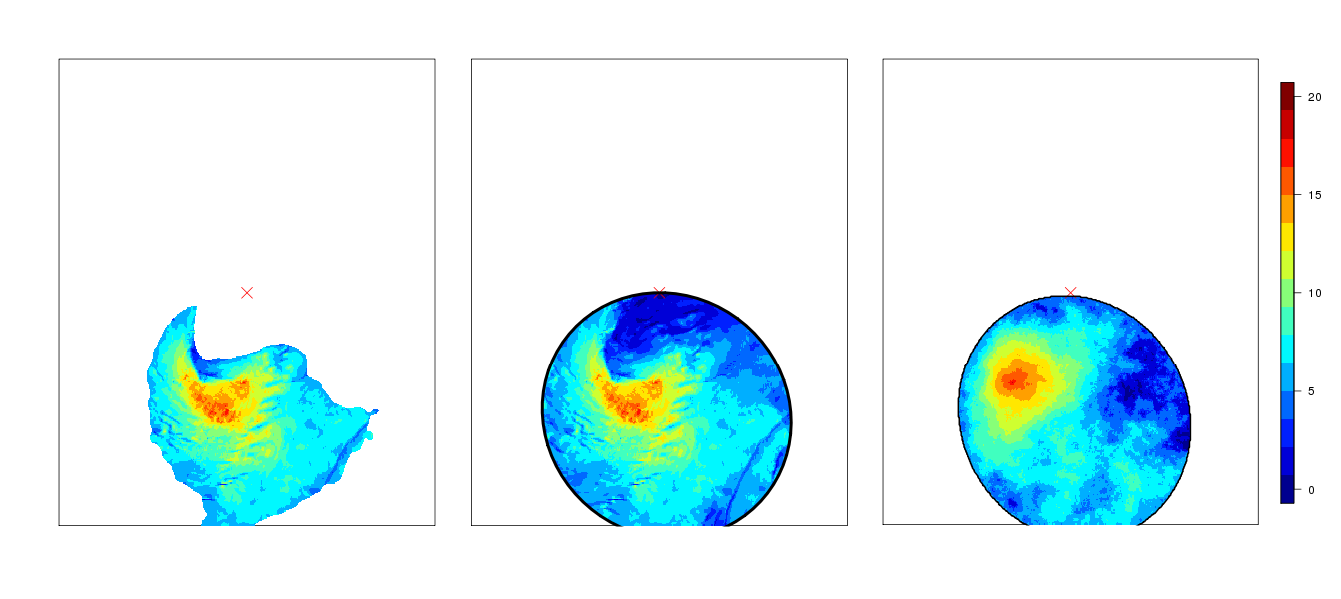}
\caption{The largest cluster of relative winds (left) after applying the spatio-temporal filter and DBSCAN clustering to the field in Figure~\ref{fig:marg_std}, the ellipse-shaped footprint determined by Khachiyan's algorithm (centre) and an example of simulating wind speeds from our model within this ellipse. (right).}
\label{fig:event_regions}
\end{figure}

\begin{figure}
\centering
\begin{tikzpicture}   
\draw[rotate=-30,line width=2pt] (0,0) ellipse (4cm and 2cm);
\draw (0,0)node[circle,fill,inner sep=2pt](a){} -- (3,1)node[label=above:$R^E_t$](r){} -- (6,2) 
node[circle,fill,inner sep=2pt,label=above:$\Omega_t$](b){};
\draw[dotted] (b) -- (6,0);
\draw (4,1.333) arc (240:270:4);
\draw (5.3,0.8)node[label=above:$\Theta^E_t$](){};
\draw (a) -- (3.464,-2);
\draw (0.3,0.7) node[label=above:$B_t$](){};
\draw (2,-1.3) node[label=above:$A_t$](){};
\draw (a) -- (1,1.732);
\draw (a) -- (-0.75,0.25)node[label=above:$R^W_t$](){} -- (-1.5,0.5)node[circle,fill,inner sep=2pt,label=left:$W_t$](){};
\draw[dotted] (a) -- (0,-2); 
\draw (0,-1) arc (270:160:1);
\draw (-0.4,-0.8) node[label=above:$\Theta^W_t$](){};
\draw[dotted] (3.464,-2) -- (3.464,0);
\draw (3.464,-0.5) arc (80:180:0.9);
\draw (2.5,-1.3) node[label=right:$\Gamma_t$](){};
\draw (6.1,2) node[label=right:$(\text{Lon}_t \text{,Lat}_t)$](){};
\draw (-3,1.5) node[label=above:$\mathcal{E}_t$](){};
\end{tikzpicture}
\caption{Graphical representation of the model variables derived from the ellipse shape used to summarise the windstorm footprint.}
\label{fig:model_par}
\end{figure}
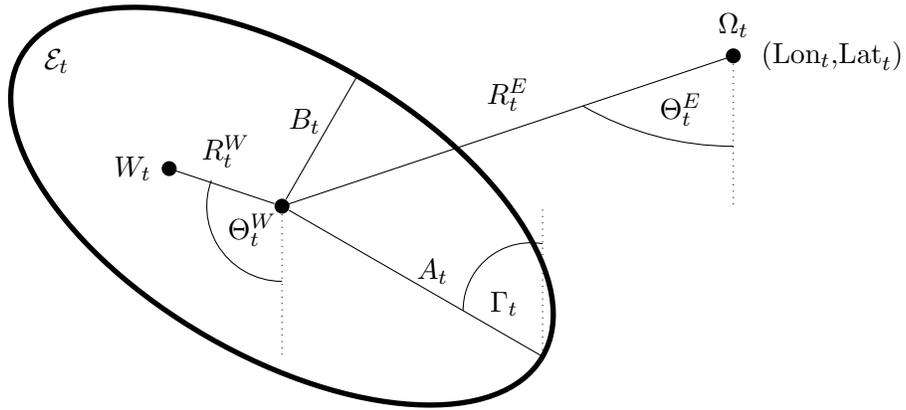

We therefore require some methodology to extract the main features of interest, such as the cluster of high relative wind speeds,
from the standardised fields and to see how these track through time. Our strategy is to identify  all large-scale high relative wind speeds associated with the extratropical cyclone at each time step, to then bound these by an elliptical region in space, and to find all such regions
associated with each cyclone track. We call each of these regions a footprint and we refer to a windstorm as being the collection of footprints over a cyclone lifetime. If a windstorm has gaps without footprints, either before, between or after periods with footprints, we refer to these as inactive and active phases of the windstorm respectively. For a storm centre $(\text{Lon}_t, \text{Lat}_t)$ with associated vorticity $\Omega_t$ at the $t$-th time step of a cyclone, we identify the footprint from the associated wind field by a multi-stage process below. To aid understanding of this process 
Figure~\ref{fig:model_par} presents the features we extract and Figure~\ref{fig:event_regions} (left and centre panels) illustrates Steps 2-4 for storm Daria. \\

\textbf{Step 1: Removal of small-scale features}\\
We apply a spatio-temporal Gaussian filter \citep{nixon2012feature} to each field, which removes the effect of the small-scale convective events.
We define a threshold $v$, indicating a large relative wind speed for the filtered data, with relative wind speeds below this level temporarily masked and those that exceed it retained. If the entire filtered wind field is masked, this is interpreted as there being no significant windstorm activity, no footprint is formed and the windstorm is deemed to be in an inactive phase. \\

{\bf Step 2: Clustering of large-scale high relative wind speeds}\\
When there is at least one exceedance of $v$ we identify clusters of exceedances using the clustering algorithm 
DBSCAN (Density-Based Spatial Clustering of Applications with Noise) \citep{ester1996density}, which recursively groups cells into distinct clusters based on adjacency to neighbouring cells, without the need to specify the number of clusters in advance. We then extract the largest of the identified clusters and define this by $S_t$ containing $m_t$ locations in 2-dimensional space such that 
$S_t=\{s_{t,1},\hdots, s_{t,m_t}\} \in \mathbb{R}^2$. \\

{\bf Step 3: Identifying of the bounding ellipse}\\
We want to model the behaviour of the wind field in the region that bounds the cluster identified in Step 2. We specify all footprints to be ellipses as we found clusters tended to have this shape and that these shapes can be described parsimoniously (by their centre, semi-major and semi-minor axes, and associated angles, see Figure~\ref{fig:model_par}). To select the ellipse to contain all elements of $S_t$ we want it to be the minimum-area ellipse containing the cluster, which is achieved by using Khachiyan's algorithm  \citep{moshtagh2005minimum,todd2007khachiyan}. A set $\mathcal{E}_t$ contained by an ellipse can be written as 
\[ 
\mathcal{E}_t = \{ s \in \mathbb{R}^2 : {(s-c_t)}^{T} E_t (s -c_t) \leq 1 \}, 
\]
where $c_t$ is the centre of the ellipse, $E_t$ is a positive-definite matrix, and the  area of $\mathcal{E}_t$ is ${\{ \det E_t^{-1} \}}^{1/2}$.
Khachiyan's algorithm finds $c_t$ and $E_t$ to minimise $\det E_t^{-1}$ subject to ${(s_{t,i} - c_t)}^{T} E_t (s_{t,i} - c_t) \leq 1$ for $i=1,\hdots,m_t$ using conditional gradient ascent methods. Once the footprint ellipse has been identified we take all the original relative wind speeds from inside the ellipse to give the footprint. At time $t$, we denote  the semi-major and semi-minor axes of the ellipse 
by $A_t$ and $B_t$ respectively, the grid cell distance between the storm centre and the centre of the ellipse by $R^{E}_{t}$, the angle between due south and the vector between the storm centre and the centre of the ellipse by $\Theta^{E}_{t} \in [-\pi,\pi]$, and the orientation of the ellipse relative to due north by $\Gamma_t \in [-\pi/2, \pi/2]$.\\

{\bf Step 4: Elimination of spurious footprints}\\
Occasionally we found footprints not generated by the extratropical cyclone spuriously identified by Steps 1-3 when
$R^{E}_{t}$ is large or ${\{ \det E_t^{-1} \}}^{1/2}$ is sufficiently small. We exclude these footprints and treat the process as in an inactive phase.\\

{\bf Step 5: Selecting features from the footprint}\\
We are mostly interested in the extreme relative wind speed and its location within footprint $\mathcal{E}_t$ with the maximum value being denoted by  $W_t$, the grid cell distance between the centre of the ellipse and the location of the maximum being $R^{W}_{t}$, and the bearing between the two cells relative to due south being $\Theta^{W}_{t} \in [-\pi,\pi]$. 
In Section~\ref{subsec:spatial} we model the entire spatial process of the relative wind speeds in footprint $\mathcal{E}_t$ conditional on the value of $W_t$ and its location. \\

\subsection{Exploratory analysis of footprint features}
\label{subsec:exploratory}
We undertake an extensive exploratory analysis on a number of aspects driving and influencing the evolution of windstorm activity, of which some findings are reported here. First, we investigate the dependence structure of characteristics of the ellipse representing the windstorm footprint, as shown in Figure~\ref{fig:model_par}. We assess the factors influencing the activation and termination of windstorm events. We also look at some quantities representing the spatial distribution of wind speeds relative to the storm centre, and how these compare with previous studies. Finally, we explore how the distribution of relative wind speeds within the footprint varies with respect to characteristics of the storm track and footprint.  \\
\begin{figure}
\centering
\includegraphics[width=10cm]{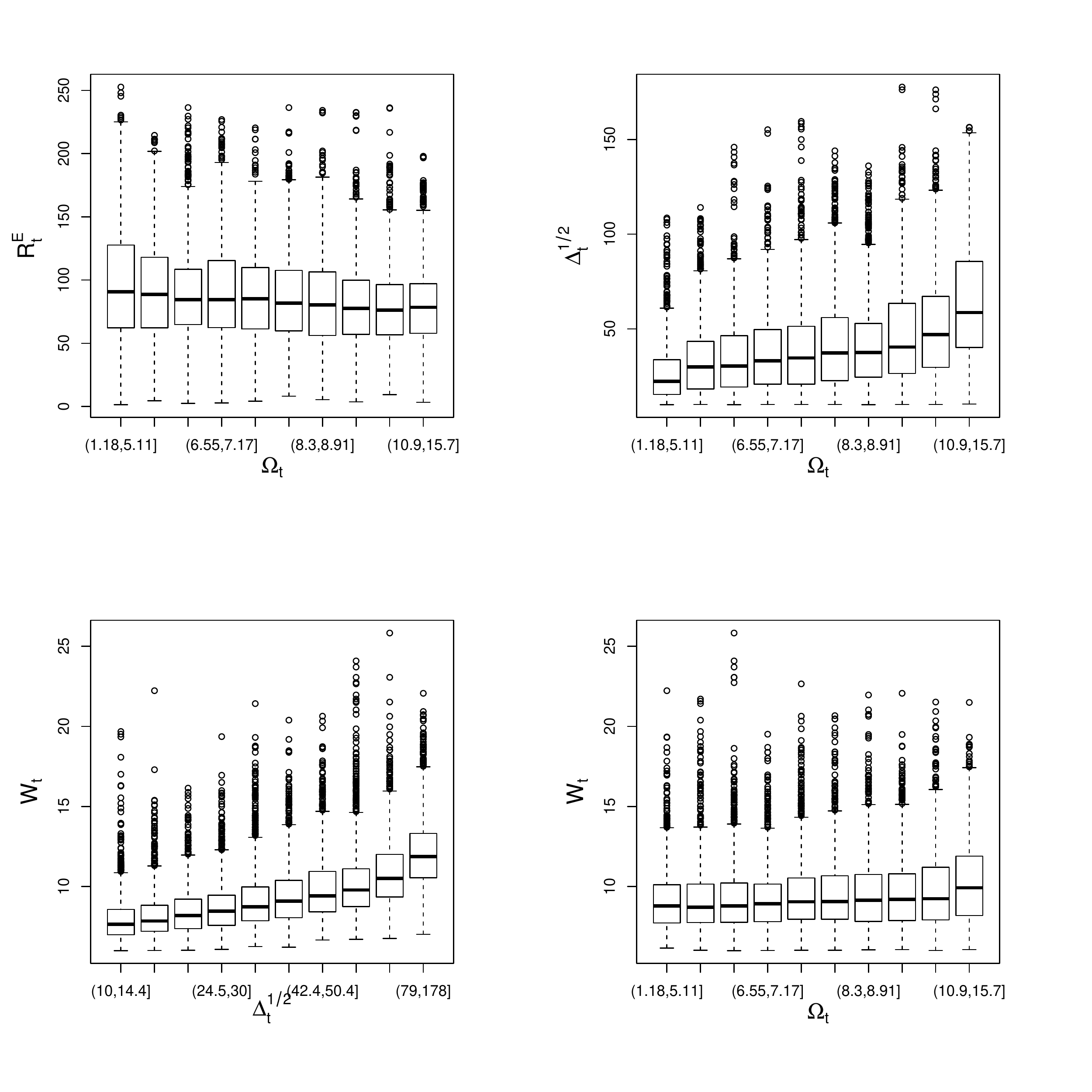}
\caption{Boxplots showing the dependence between $\Omega_t$ and $R^{E}_t$, $\Omega_t$ and $\Delta^{1/2}_t$, $\Delta_t^{1/2}$ and $W_t$, and $\Omega_t$ and $W_t$. }
\label{fig:dependence}
\end{figure}

Figure~\ref{fig:dependence} shows boxplots illustrating some key dependencies between variables of a footprint. The area of the ellipse, which is proportional to $\Delta_t = A_t \times B_t$, tends to increase as vorticity and maximum relative wind speed increases, indicating that the strongest events tend to occur on a larger spatial scale. The radius $R^{E}_t$ tends to decrease as $\Omega_t$ increases, although the effect is small, suggesting that footprints tend to occur closer to the storm centre when a large vorticity is observed. Maximum relative wind speed $W_t$ tends to increase as $\Omega_t$ increases, though this dependence is weak. We also examine partial autocorrelation plots (not shown) to determine how individual components of the ellipse depend on their lags. This analysis shows evidence of a second-order temporal dependence structure in most components reflecting the smooth evolution of footprints through the windstorm. \\

\begin{figure}[h!]
\centering
\includegraphics[width=7cm]{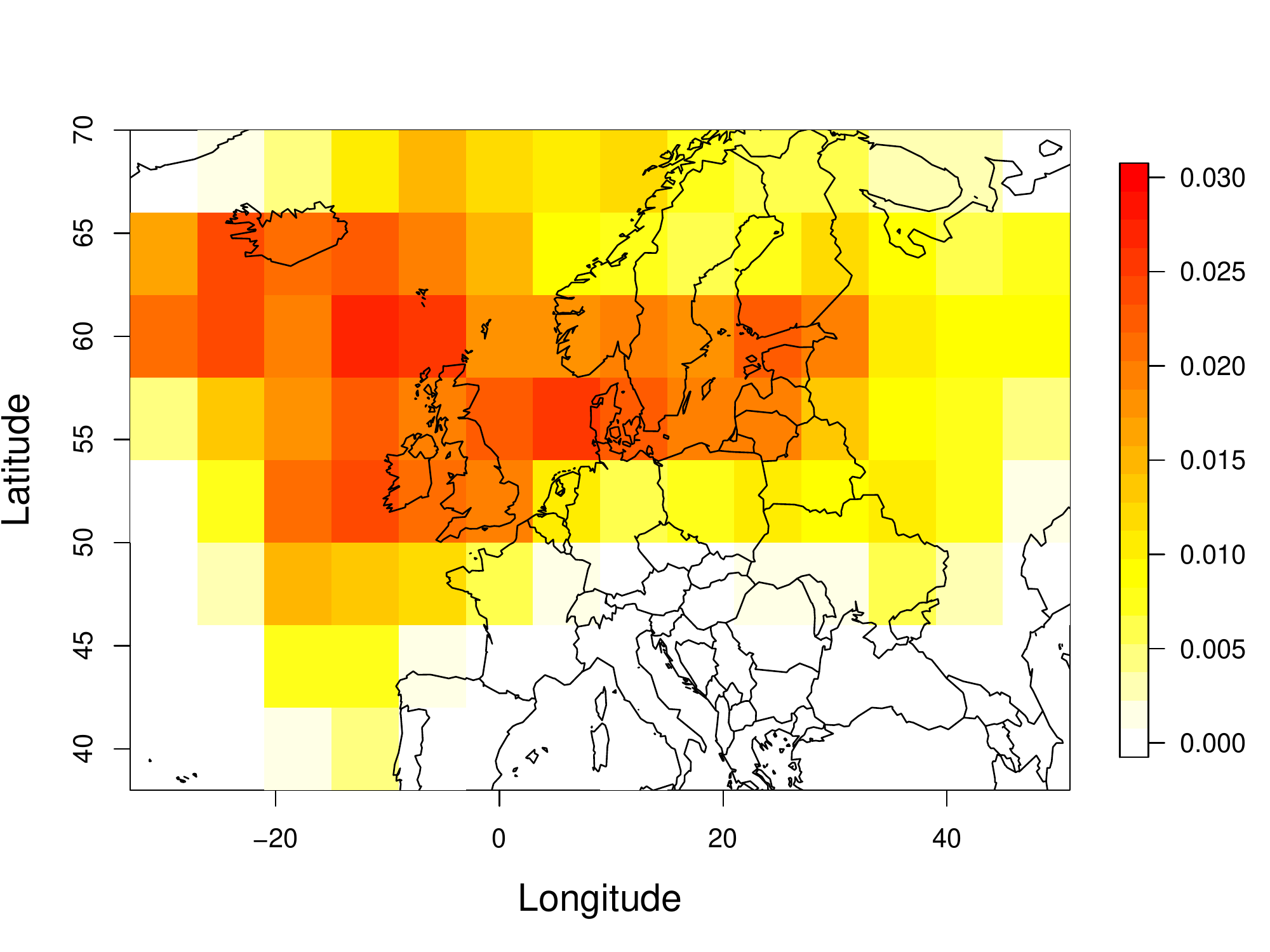}
\includegraphics[width=7cm]{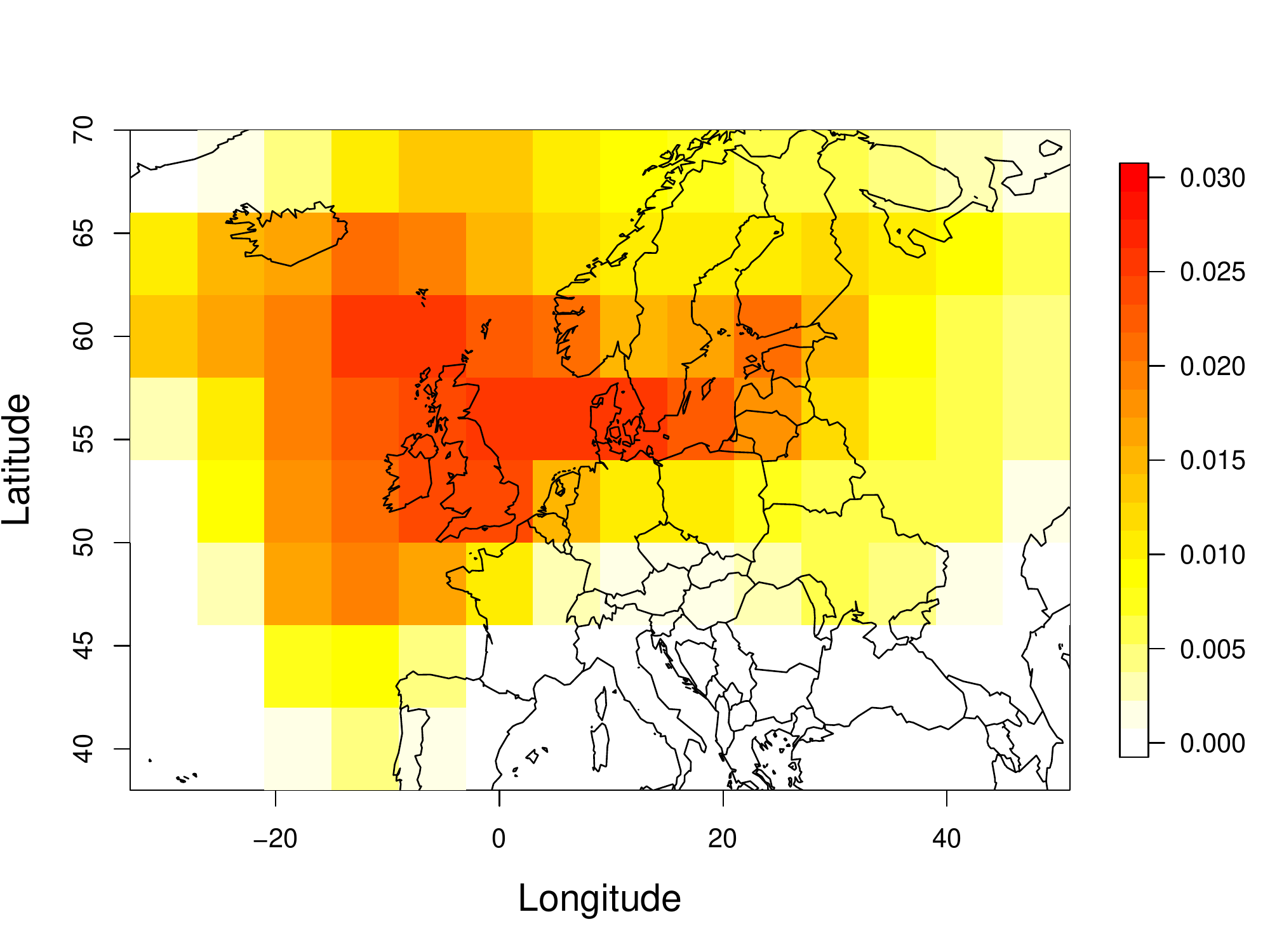}
\caption{The spatial density of track locations that are associated with observed (left) and simulated (right) windstorms.}
\label{fig:storm_dens_sim}
\end{figure}
Next consider what factors influence whether a windstorm is active or not. With regards to windstorm activation, we investigate the components of the track that may trigger an event. We find that the probability of windstorm activations tends to increase as vorticity increases, signalling a direct link between the intensity of the storm track and the occurrence of windstorm events. With regard to an inactive phase caused by termination of an active phase, we need to account for the variables of the windstorm footprint in addition to information from the track. We find that a storm is more likely to terminate if it is associated with small values of relative maximum wind speed, vorticity and area. These findings indicate a termination is more likely if the windstorm weakens both in terms of its magnitude and its spatial scale. \\

This exploratory analysis also shows some spatial variation in the occurrence of footprints. Figure~\ref{fig:storm_dens_sim} (top panel) shows the density of storm track locations when windstorms are in an active phase. This figure shows that windstorms tend to occur over the North Atlantic and western Europe, with the density decaying as one moves to the edges of the domain. The reductions in density on the western boundary are the result of the filtering of footprints and the boundary effect on the EURO4 simulation of weather. Most extratropical cyclones enter the EURO4 domain through this western boundary but their intensity is diminished due to the coarser resolution of the driving GCM. Their subsequent intensification takes a number of time steps which results in smaller footprints and lower winds towards this boundary (see Figure~\ref{fig:gpd_par}). \\

\begin{figure}
\centering
\includegraphics[width=14cm]{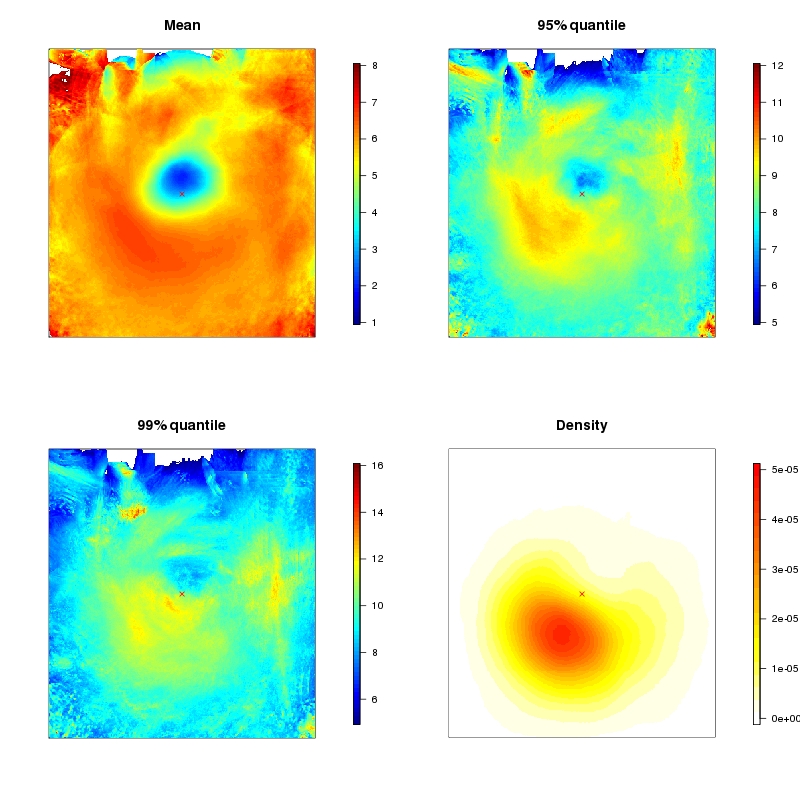}
\caption{The mean, $95\%$ quantile, $99\%$ quantile and density of wind speeds relative to the storm centre (represented here as a cross) over all observed footprints.}
\label{fig:footprint_stats}
\end{figure}
For each windstorm in an active phase, i.e., with an identified footprint, we collect all fields centred on the storm's maximum vorticity, like in Figure~\ref{fig:event_regions}, and assess the spatial distribution of these fields relative to the storm centre.  Figure~\ref{fig:footprint_stats} shows the mean, $95\%$ and $99\%$ quantiles of wind speeds. This figure shows that relative wind speeds tend to be larger in regions southwest of the storm centre, which arise with the passage of cold fronts. This figure also shows the density of events over all footprints, illustrating that windstorms are most likely to occur southwest of the storm centre, with very few events occurring in the northern half of the field in comparison. These are consistent with the observations of \citet{catto2010} and \citet{rudeva2011}. \\

Unusual events with large magnitudes are detected on the northwest and southeast edges of the domain. Despite our filtering, these events may not be generated by the extratropical cyclone; however, they are very rare and their impact minimal. The mean behaviour shows a local minimum occurring close to the storm centre, which likely arises as a result of low wind speeds occurring at pressure minima \citep{catto2010}, which is known to be spatially adjacent to where the maximum vorticity occurs \citep{hoskins2002new}. \\ 

We investigate the factors influencing the distribution of relative wind speeds within each footprint through analysis of their mean and standard deviation. This reveals, intuitively, that the mean and standard deviation of winds tend to increase as the maximum relative wind increases. In contrast, an increase in vorticity tends to slightly reduce the mean relative wind, while increasing the standard deviation. Additionally, the standard deviation of wind increases as the area of the ellipse increases, which one would expect given the increased chances of observing smaller and larger values of wind speed. By construction the strongest relative winds tend to occur near the location of the maximum, determined by $R^{W}_t$ and $\Omega^{W}_t$, while weaker relative winds are more likely closer to the perimeter of the ellipse. Additionally, the relative winds tend to exhibit anisotropic properties, which tends to manifest in a `stretching' of the band of strongest winds oriented in the direction perpendicular to $\Theta_t$. As the windstorm evolves, this gives the effect of the winds `bending' around the storm centre.

\section{Windstorm modelling}
\label{sec:methodology}

\subsection{Introduction}
\label{subsec:intro}
We propose an approach for modelling and simulating windstorms relative to an extratropical cyclone track that is motivated by our findings in Section~\ref{subsec:exploratory}. In particular, along a cyclone track we model whether the windstorm is active or not at each time step, and when it is active we model the evolution of the footprint's characteristics along with the spatial distribution of wind speeds within a footprint. 
Specifically if $[t_{S},t_{T}]$ is a contiguous active interval, with inactive periods immediately before and afterwards, 
we call $t_S$ and $t_T$ the activation and termination times respectively. Windstorm can have repeated phases of activity and inactivity along a cyclone track. For each active phase the features of the process identified in Figure~\ref{fig:model_par}, denoted  $\boldsymbol{Z}_t$ at time $t$, are jointly modelled over time for all $t_{S}\le t \le t_{T}$ in Section~\ref{subsec:footprint}; whilst we present models for the occurrence $t_S$ and $t_T$ in Section~\ref{subsec:activate}, and in Section~\ref{subsec:spatial} we model the wind speed fields within the footprint.


\subsection{Footprint modelling}
\label{subsec:footprint}
During active phases, we model the joint temporal evolution of the footprints variables $\boldsymbol{Z}_t =\{A_t,B_t,W_t,R^{E}_t,\Theta^{E}_t, R^{W}_t, \Theta^{W}_t, \Gamma_t \}$ of Figure~\ref{fig:model_par} conditional on the storm track variables
$(\text{Lon}_t,\text{Lat}_t, \Omega_t)$. For simplification purposes for our inference, in practice we condition on where 
$(\text{Lon}_t, \text{Lat}_t)$ in $\nabla_t$, where $\nabla_t$ is a region of size $20^\circ \times 14^\circ$ centred at $(\text{Lon}_t,\text{Lat}_t)$. The exploratory analysis of  Section~\ref{subsec:exploratory} motivates modelling $\boldsymbol{Z}_t$ as 
a $k$-th order Markov process during active phases along the cyclone track and for $\boldsymbol{Z}_{t}$ and $\boldsymbol{Z}_{t^\prime}$ to be conditionally independent of each other given the track information if $t$ and $t^\prime$ correspond to different cyclone tracks or different 
active phases along a single track. \\

By the Markov property, the distribution of the current value of a process is affected only by the previous $k$ time steps of the process.  Let $\boldsymbol{Z}_{i:j} = \{\boldsymbol{Z}_{t}: t = i,\hdots,j\}$, where $i\le j$, then it is only necessary to model the joint distribution of $\boldsymbol{Z}_{t:t+k}$, from which the conditional density function of $\boldsymbol{Z}_{t+k} \mid \boldsymbol{Z}_{t:t+k-1}$ can be derived. 
Empirical evidence, such as Figure~\ref{fig:dependence}, helps to identify which components of $\boldsymbol{Z}_{t:t+k}$ are independent or conditionally independent, which helps to identify $k$.  For example, we found $k=2$ and a weakly dependent relationship between $\Omega_t$ and $W_t$ with $\Omega_t$ and $W_t$ conditionally independent given $A_t$ and $B_t$. This result allows us to simplify our model for $\Pr(\boldsymbol{Z}_{t+k} \leq z \mid \boldsymbol{Z}_{t:t+k-1} = \boldsymbol{z}_{t:t+k-1})$. First, consider an initialisation time $t_A$, with
$t_S\le t_A\le t_T$, which is determined by an algorithm in Section~\ref{subsec:activate}. We simulate the initialisation value $\boldsymbol{Z}_{t_A}$ from the distribution of $\boldsymbol{Z}_{t_A} \mid (\text{Lon}_{t_A}, \text{Lat}_{t_A}, \Omega_{t_A})$, which we estimate using a conditional kernel density (see Appendix~\ref{App:sim_kernel}). Then consider
forward propagation of $\boldsymbol{Z}_{j}$, given $\boldsymbol{Z}_{t_A}$, for $t_A< j \le t_T$, using the following conditional distributions: 
\begin{eqnarray*}
R^{E}_j  &\mid & R^{E}_{{(j-k)}_{\geq t_A}:j-1} = r^{E}_{{(j-k)}_{\geq t_A}:j-1}, \Omega_j = \omega_j,  \nonumber \\
\Theta^{E}_j &\mid & \Theta^{E}_{{(j-k)}_{\geq t_A}:j-1}= \theta^{E}_{{(j-k)}_{\geq t_A}:j-1}, R^{E}_j = r^{E}_j;  \nonumber\\
A_j & \mid &A_{{(j-k)}_{\geq t_A}:j-1}= a_{{(j-k)}_{\geq t_A}:j-1}, \Omega_j = \omega_j, R^{E}_j = r^{E}_j, \Theta^{E}_j = \theta^{E}_j;  \nonumber\\
B_j &\mid &B_{{(j-k)}_{\geq t_A}:j-1}= b_{{(j-k)}_{\geq t_A}:j-1}, \Omega_j = \omega_j, R^{E}_j = r^{E}_j, \Theta^{E}_j = \theta^{E}_j; \nonumber\\
W_j &\mid & W_{{(j-k)}_{\geq t_A}:j-1}= w_{{(j-k)}_{\geq t_A}:j-1}, R^{E}_j = r^{E}_j, \Theta^{E}_j = \theta^{E}_j, A_j = a_j, B_j = b_j;  \nonumber\\
\Gamma_j & \mid &\Gamma_{{(j-k)}_{\geq t_A}:j-1}= \gamma_{{(j-k)}_{\geq t_A}:j-1}, \Theta^{E}_j = \theta^{E}_j; \nonumber\\
R^{W}_j & \mid & R^{W}_{{(j-k)}_{\geq t_A}:j-1}= r^{W}_{{(j-k)}_{\geq t_A}:j-1}, A_j = a_j;  \nonumber\\
\Theta^{W}_j & \mid & \Theta^{W}_{{(j-k)}_{\geq t_A}:j-1}= \theta^{W}_{{(j-k)}_{\geq t_A}:j-1}, \nonumber
\end{eqnarray*}
where we use the notation ${(j-k)}_{\geq t_A} = \max\{ j-k, t_A\}$ and in each case the distribution is conditional on $(\text{Lon}_j,\text{Lat}_j)$ being in $\nabla_j$. Realisations for $R_{j}^W$ and $\Theta_{j}^{W}$ are rejected if the simulated position of the maximum occurs outside of the footprint at time $j$.  We model each of these conditional distributions using 
kernel density estimates, the formulation of which can be found in Appendix~\ref{App:sim_kernel}. 
Backwards simulation for $t_T\le j<t_A$ is implemented similarly, but with substituting $\boldsymbol{Z}_{j}$ for $\boldsymbol{Z}_{t_A -j}$.

\subsection{Initialisation, activation and termination}
\label{subsec:activate}
 
We have to model the probability distributions for initialisation time $t_A$, activation time $t_S$ and termination $t_T$, for each active period given other covariates of the footprint and storm track process. The model formulation is identical for each of these, with $t_A$ being generated first and $t_S$ and $t_T$ conditional on $t_A$. 
Let $T_{t}$ be a Bernoulli random variable such that: $T_{t} = 1$ if  the storm is active at time $t$ and $T_{t} = 0$ otherwise.
We model $T_t$ by a Bernoulli logistic generalised additive model \citep{wood2006generalized}. 
So $T_{t} \sim$ Bernoulli$(p_{t})$, where
\[ p_t = \frac{\exp\left\{ \sum_{i=1}^{q} \beta_{i} (\nu_{i,t}) \right\}}{1+\exp\left\{\sum_{i=1}^{q} \beta_{i} (\nu_{i,t}) \right\}} ,
\]
where $\beta_{i}$ is a smooth non-linear function of covariate $\nu_{i}$ with $i \in (1,\hdots,q)$, where $q$ is the number of covariates and $\nu_{i,t}$ denotes the realisation of the $i$th covariate at time $t$. The smooth functions are represented by penalised regression splines, where the smoothing parameter is determined using generalised cross validation (GCV) and the model is fitted using penalised maximum likelihood,
with the choice of covariates we used discussed in Section~\ref{subsec:val_foot_model}. \\

Now consider how to determine the initialisation time $t_A$ out of the possible values of $1\le t \le \ell$, where $\ell$ the duration of the storm track. Section~\ref{subsec:exploratory} demonstrated that the windstorm is typically in an active phase at the time when the maximum vorticity is observed, as this time is associated with the strongest winds over the cyclone track and hence has the highest
probability $p_t$ of being active. We denote this time by $t_{\Omega}$, with $1\le t_{\Omega} \le \ell$,  
If the windstorm is determined to be active at $t_{\Omega}$, i.e., $T_{t_{\Omega}}=1$, then we set $t_A=t_{\Omega}$.
If a windstorm is inactive at $t_{\Omega}$, i.e., $T_{t_{\Omega}}=0$, we attempt to initialise the active phase successively forwards, and then backwards, in time until an active phase is first identified, i.e., the first occasion when we get a Bernoulli realisation of 1. If the initialisation is found during the forwards procession then $t_A > t_{\Omega}$ and $t_S = t_A$ and we only propagate the Markov chain forwards. Similarly, initialisation on the backwards procession gives $t_A < t_{\Omega}$ we set $t_T = t_A$. We allow for the possibility that multiple phases of consecutive footprints can occur on the same track. If $t_{S} > 1$, we proceed backwards along the track to check if the windstorm reactivates, in which the same procedure applies until $t=1$. Similarly, if $t_{T} < \ell$, we proceed forwards along the track to check if the windstorm reactivates, in which the same procedure applies until $t=\ell$. \\ 

\subsection{Modelling wind speeds within a footprint}
\label{subsec:spatial}
We also require an approach for modelling the spatial distribution of relative wind speeds within a footprint at each time step of a windstorm.  In theory many different spatial processes could be used, but a natural class of models to consider is Gaussian processes, which are widely used in geostatistics \citep{cressie1993statistics, stein1999interpolation, diggle2007geo} to model spatial data as they are simple to use and parsimonious. A Gaussian process describes the joint distribution of random variables over a continuous domain such as space inside a footprint, while for any finite collection of locations in the space the variables follow a multivariate Gaussian distribution. \\

Let $\{X^E (s,t) : s \in \mathcal{E}_t\}$ denote the field of relative wind speeds in the ellipse $\mathcal{E}_t$ at time $t$, where $X^{E} (s,t)$ is marginally Exp$(1)$ distributed over $t$ for each $s$. To model using Gaussian processes we first need to transform the variables marginally to be Gaussian. Let $D_t$ be the distribution function of $X^E (s,t)$ for all $s \in \mathcal{E}_t$ at time $t$. As the values in the footprint are not typical of the wind field, $D_t$ is not the distribution function of an Exp$(1)$ variable. Based on findings from  Section~\ref{subsec:exploratory}, we use a weighted nonparametric estimate $\tilde{D}_t$ of the distribution function $D_t$. Using a kernel, we weight observed relative wind speeds in $\mathcal{E}_t$ conditional on the corresponding values of $W_t$, $\Delta_t$ and $\Omega_t$. We use a probability integral transform to convert to a Gaussian field, which we denote by $X^{G}(s,t)$, with
\begin{equation} 
X^{G} (s,t) = \Phi^{-1} \left[ D_t (X^{E}(s,t))\right],
\label{eq:gauss_trans}
\end{equation}
for all $s \in \mathcal{E}_t$ and all $t$, where $\Phi$ denotes the standard Gaussian distribution function.\\

We make the assumption that 
for each $t$, $\{X^{G}(s,t) : s \in \mathcal{E}_t \}$ follows a Gaussian process with zero mean and 
unit variance. The correlation  between 
sites $(s_i,s_j)\in \mathcal{E}_t$ is modelled by $\rho ( (s_i,s_j) J_t \Psi_t)$ where $\rho (\cdot)$ denotes an 
isotropic correlation function and the time varying anisotropic effects are described by the matrices
\[ \Psi_t =
\begin{bmatrix}
\cos \psi_t & -\sin \psi_t \\
\sin \psi_t & \cos \psi_t \\
\end{bmatrix}
\hspace{20pt}
J_t =
\begin{bmatrix}
1 &  0 \\
0 & \zeta_t \\
\end{bmatrix},
\]
where  $\psi_t$ is the time-varying anisotropy angle representing the counter-clockwise rotation of the space, given by
and $\zeta_t > 1$ is the time-varying anisotropy ratio, which controls the degree of stretching along the angle where correlation decays most slowly with increasing distance. Supported by the analysis of within-footprint winds and their anisotropic properties, 
we fix $\psi_t$ to be the angle perpendicular to $\Theta_t$ and set $\zeta_t=A_t/B_t$. \\

The correlation function is typically chosen so that the correlation between $X^{G}(s_1,t)$ and $X^G (s_2,t)$ decreases as the distance $|s_2-s_1|$ between the sites increases. A common choice of correlation function is the Mat{\'e}rn family, which has the form
\begin{equation}
 \rho(u) = {\{ 2^{\kappa -1} \Gamma(\kappa)\}}^{-1} {(u/\alpha_t)}^{\kappa} K_{\kappa} (u/\alpha_t),
 \label{eq:matern}
 \end{equation}   
where $\kappa > 0$ is a shape parameter that determines the smoothness of the underlying process, $K_{\kappa}$ denotes a modified Bessel function of the second kind of order $\kappa$, and $\alpha_t > 0$ is a time-varying scale parameter, with dimensions of distance, with increasing $\alpha_t$ corresponding to stronger correlations. The Mat{\'e}rn family is a generalisation of other common choices of correlation functions including the exponential ($\kappa =0.5$) and Gaussian ($\kappa \rightarrow \infty$), with larger $\kappa$ giving smoother fields. We fix $\kappa = 0.6$, which gives fields of similar smoothness to the observed fields. We estimate the parameter $\alpha_t$ for each footprint using variogram methods. We avoid the use of likelihood methods due to the computational difficulties that arise with large spatial datasets like ours. Our investigations (not shown here) suggest that $\alpha_t$ is constrained by the value of $\Delta_t$, which is proportional to the area of the ellipse. We therefore model $\alpha_t \mid \Delta_t $ using a conditional kernel density, the formulation of which is outlined in Appendix~\ref{App:sim_kernel}. At time $t$, we generate a realisation of $\alpha_t$ conditional on the simulated realisation of $\Delta_t$ determined by the model in Section~\ref{subsec:footprint}. \\

We incorporate information about the physical structure of the footprint in determining the structure of the Gaussian field $\{\tilde{X}^G (s,t) : s \in \mathcal{E}_t\}$ by using conditional simulation \citep{diggle2007geo}. We impose three conditions on the simulated fields: that the maximum relative wind speed is simulated at the position determined by $R^{W}_t$ and $\Theta^{W}_t$; that the lower limit of $\tilde{D}_t$ occurs on the outer perimeter of the ellipse; and that the lower limit of $\tilde{D}_t$ occurs everywhere in the region corresponding to the local minimum of wind speeds (see Section~\ref{subsec:exploratory}) near the storm centre when a footprint is simulated in the vicinity of this region. The first and second conditions create a two-dimensional pseudo-Brownian bridge between the position of maximum and the perimeter of the footprint. To impose the third condition, we specify a second ellipse centred at the local minimum with random dimensions for size; specifically we fix the maximum length of the semi-major and semi-minor axes of this ellipse to be 40 and 35 units of grid-cell distance respectively, with random perturbations modelled as Exp(0.05) random variables. These values were found to replicate well the average behaviour of wind speeds in the region at which the local minimum occurs. \\

After simulation of $\{\tilde{X}^{G} (s,t): s \in \mathcal{E}_t \}$, we transform this field to obtain a field of relative wind speeds conditional on the characteristics of the footprint such that
\[ \tilde{X}^E (s,t) = \tilde{D}_t^{-1} \left(\Phi[\tilde{X}^G(s,t)] \right). \]
An example of this is shown in Figure~\ref{fig:event_regions}, in which we have simulated a field of relative wind speeds conditional on the footprint of storm Daria at this particular time step. Our model captures the correlation structure of the field quite well, with the weakest winds occurring on the outer perimeter of the ellipse and the large winds occurring in similar locations to the observed footprint. For this simulated field,  the decay from the maximum relative wind speed appears more isotropic than the observed field. When viewed in an Eulerian framework, this level of spatial difference is not too important as these footprints move over space with the cyclone track, so blur out this distinction.
Having obtained the simulated relative wind fields, we can then transform these onto the observed margins, such that for each $s \in \mathcal{E}_t$
\[ \tilde{X}(s,t) = F_{s}^{-1} (\tilde{X}^E(s,t)), \]
where $F_{s}$ denotes the marginal model for cell 
$s$, as defined in~(\ref{eq:marginal_model}). With this formulation, we are assuming the relative wind fields at consecutive time steps are conditionally independent given temporally correlated realisations of $W_t$ and $\alpha_t$. While this assumption gives good results in practice, further investigation may be necessary to assess whether performance could be improved by specifying a spatio-temporal structure in the correlation function $\rho(\cdot)$.

\section{Results}
\label{sec:results}
We examine the performance of the windstorm model in terms of simulated footprint characteristics and then the wind speeds within the footprint. The joint risk from extreme windstorms at locations in northern England and eastern Germany is then explored by combining the windstorm model presented here with the track model of \citet{sharkey2017tracking} to produce estimates of joint event probabilities through simulation.

\subsection{Validation of footprint model}
\label{subsec:val_foot_model}
\begin{figure}
\centering
\includegraphics[width=10cm]{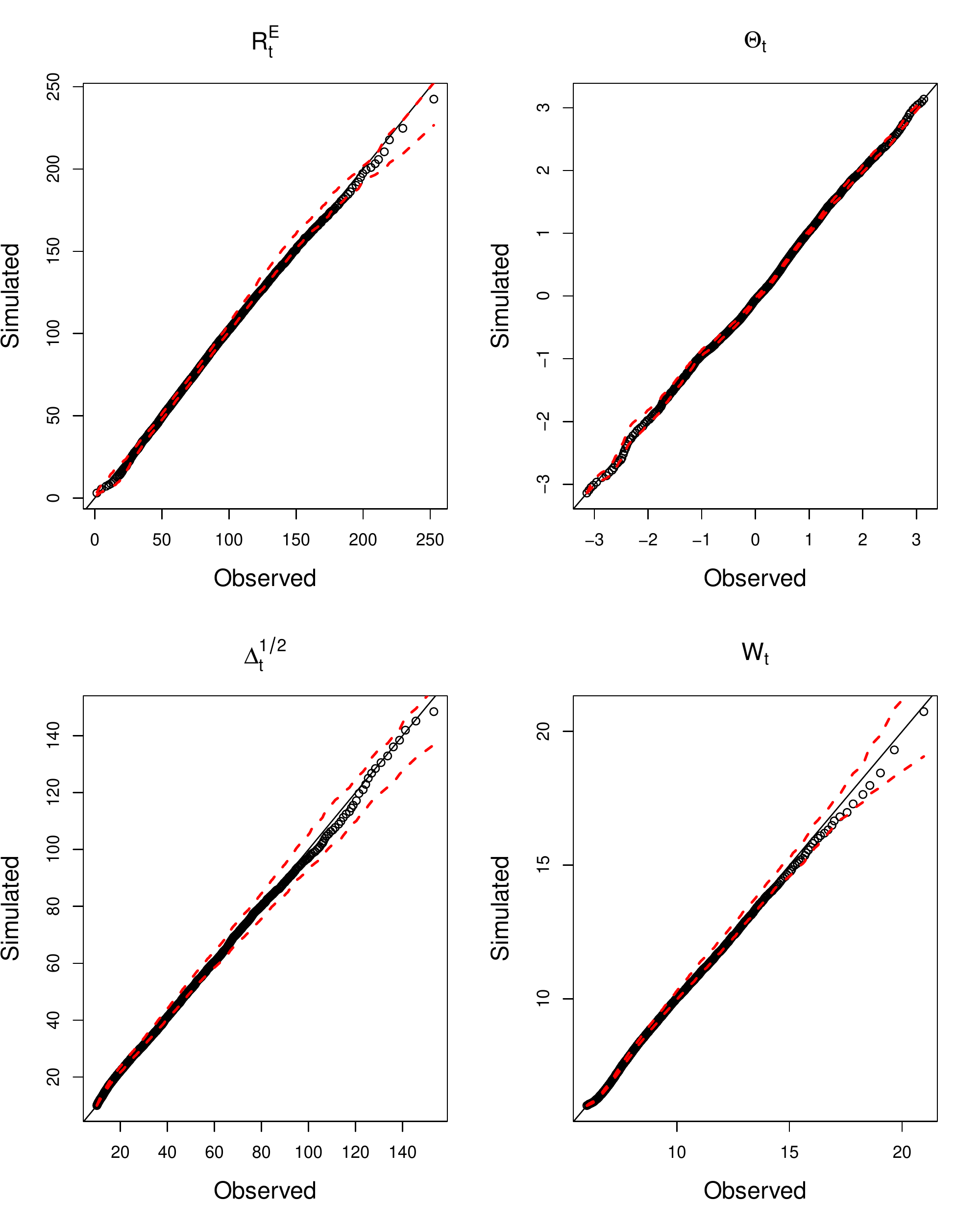}
\caption{QQ plots, with 95$\%$ tolerance intervals, comparing the observed and simulated marginal distributions of $R^{E}_t$, $\Theta_t$, $\Delta_t^{1/2}$ and $W_t$. Simulated values are based on footprints relative to 2,944 synthetic tracks from the model of \citet{sharkey2017tracking}.}
\label{fig:qq_footprint}
\end{figure}
\begin{figure}[h!]
\centering
\includegraphics[width=12cm]{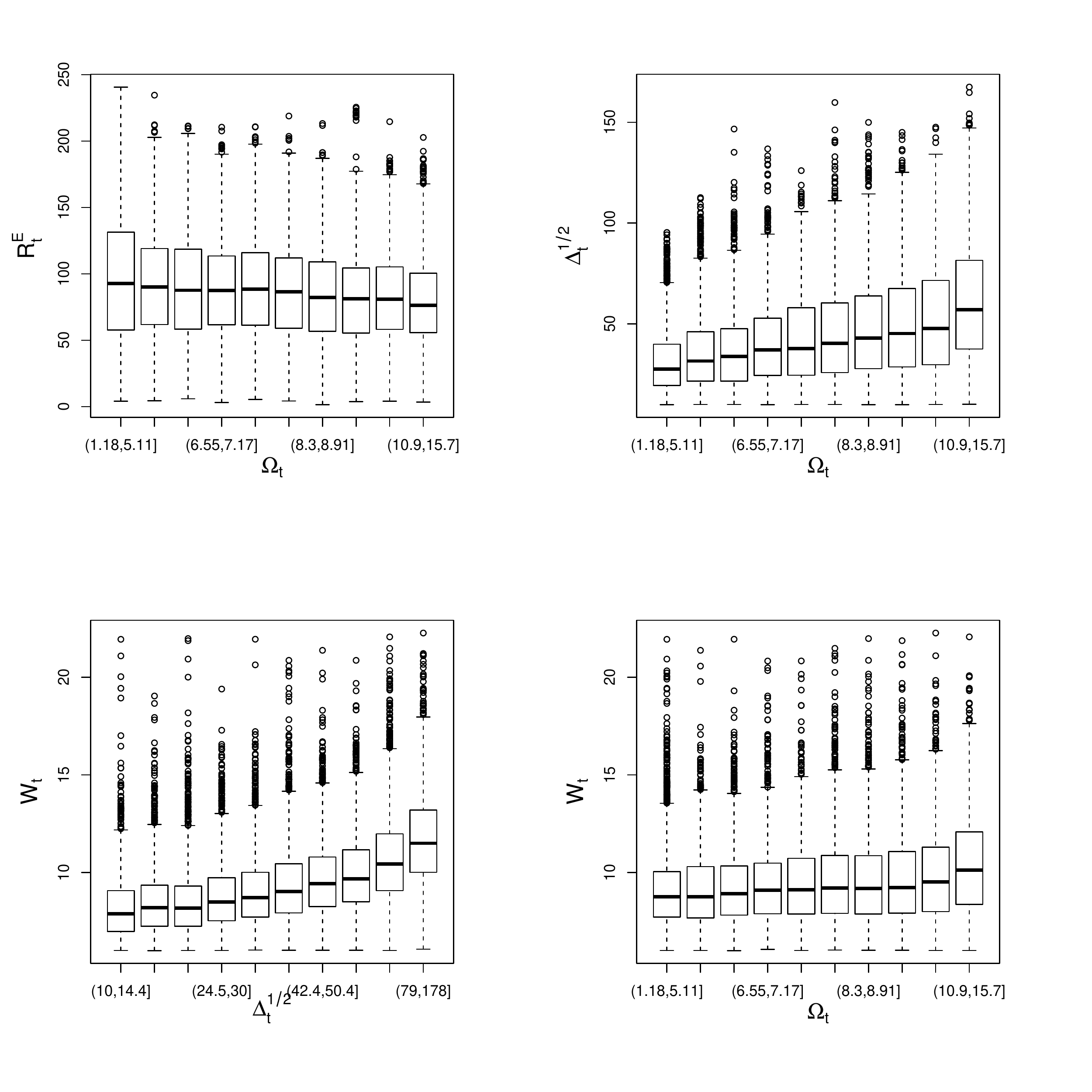}
\caption{Boxplots showing dependence structure of some aspects of the simulated footprints. Simulated values are based on footprints relative to 2,944 synthetic tracks from the model of \citet{sharkey2017tracking}.}
\label{fig:dependence_sim}
\end{figure}

We explore first whether the characteristics of windstorm footprints are being captured through an assessment of the marginal distributions of the individual components and their dependence structure. QQ plots based on the simulation of footprints using the model described in Section~\ref{subsec:footprint}, applied to $2,944$ synthetic storm tracks, the same number as in the observed record, were assessed both for the observed tracks and tracks generated by the model of \citet{sharkey2017tracking}. Both showed similar positive results, so we illustrate only the latter (see Figure~\ref{fig:qq_footprint}). They show that the marginal distributions of radius, bearing, proportional area and relative maximum wind speed are being captured well by the model. Figure~\ref{fig:dependence_sim} shows that, when compared with Figure~\ref{fig:dependence}, the dependence structure of these components is also consistent with the observations. We can thus conclude that the physical structure of the observed windstorm footprints is replicated sufficiently by the model.  \\

\begin{figure}[h!]
\centering
\includegraphics[width=15cm]{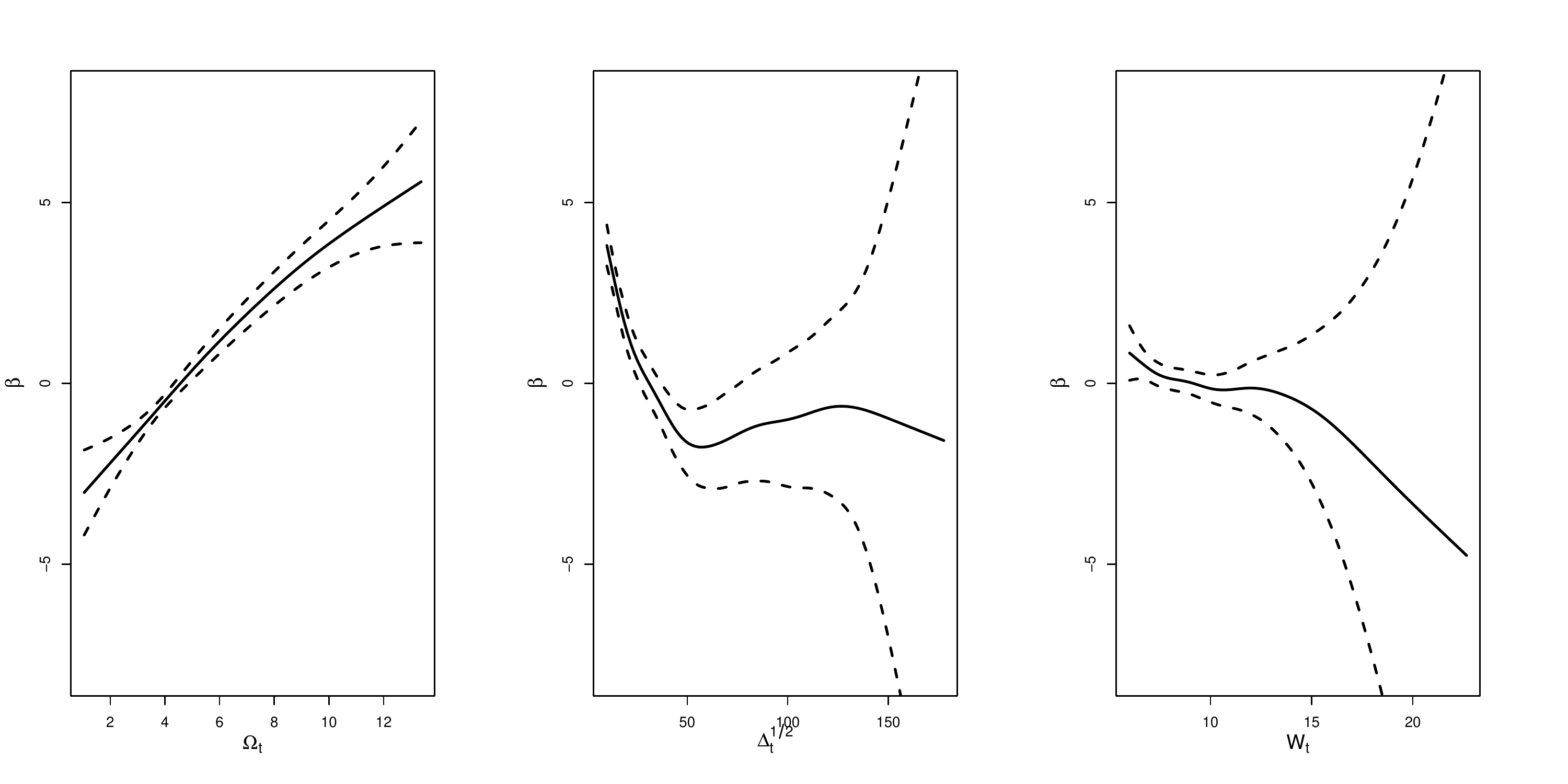}
\caption{The smooth functions $\beta_i$ (with 95$\%$ confidence intervals) showing the effect of $\Omega_t$ on the probability of windstorm activation (left) and of $\Delta_t^{1/2}$ (centre) and $W_t$ (right) on the probability of termination.}
\label{fig:GAM_outputs}
\end{figure}
We examined the components of a windstorm influencing the activation and termination models that were outlined in Section~\ref{subsec:activate}. In both cases, we investigate multiple combinations of covariates and compare model fit using AIC. The best fitting activation model includes functions of vorticity, longitude and latitude. Figure~\ref{fig:GAM_outputs} (left panel) shows the estimated smooth function $\beta_i$ associated with vorticity, which shows that $\beta_i$ tends to increase approximately linearly as vorticity increases. This relationship has the effect that the probability of activation tends to increase as vorticity increases, which reflects our findings from Section~\ref{subsec:exploratory}. The best fitting termination model includes functions of $\Delta^{1/2}_t$ and $W_t$. Figure~\ref{fig:GAM_outputs} (centre and right panels) shows that $\beta_i$ tends to decrease non-linearly as both $\Delta_t^{1/2}$ and $W_t$ increase, which means that the probability of termination also tends to decrease. The effect of $\Delta_t^{1/2}$ tends to level off at high values, though wide confidence intervals suggest that this effect is highly uncertain. This analysis confirms our belief that weakening events in terms of spatial extent and maximum relative wind speed are consistent with the termination of an active phase of a windstorm. Figure~\ref{fig:storm_dens_sim} (bottom panel) shows the spatial density of windstorm occurrence in the simulations. We see that the large-scale spatial variation of windstorms from the model reflects that of the observations in Figure~\ref{fig:storm_dens_sim} (top panel).

\subsection{Validation of model for wind speeds within a footprint}

\begin{figure}[h]
\centering
\includegraphics[width=14cm]{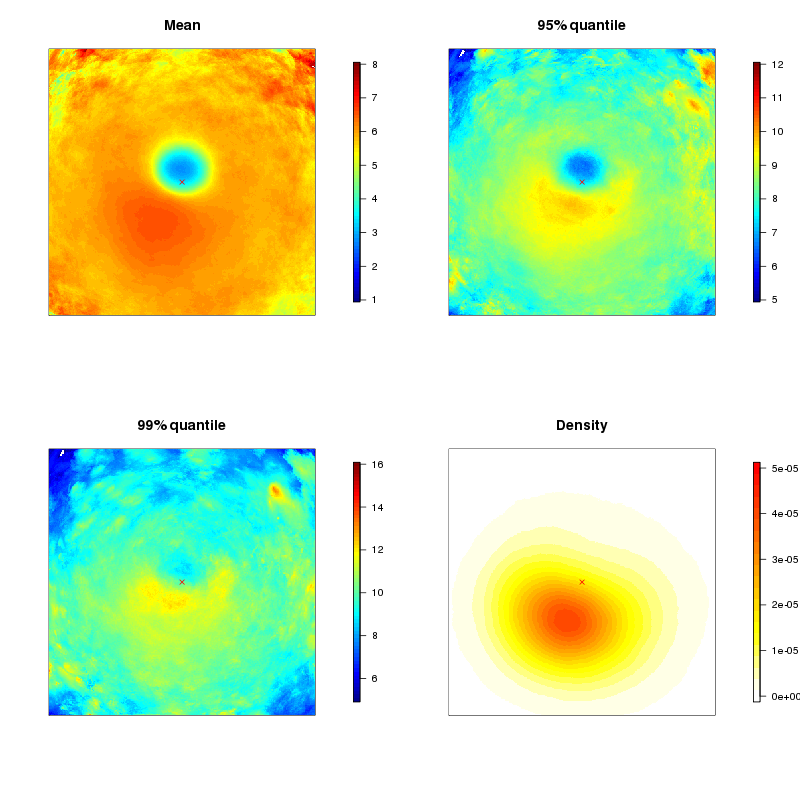}
\caption{The mean, $95\%$ quantile, $99\%$ quantile wind speed and density of within-footprint winds relative to the storm centre (represented by the cross) over a set of simulated windstorms.}
\label{fig:spatial_sim}
\end{figure}

We check that the model replicates well the physical structure and location of winds relative to the storm centre. For this task we simulate the wind speeds within the footprints generated in Section~\ref{subsec:val_foot_model} using the wind model of Section~\ref{subsec:spatial}. Figure~\ref{fig:spatial_sim} shows the mean, $95\%$ quantile, $99\%$ quantile and spatial density of simulated winds from all simulated footprints relative to the centre of the storm, quantities that were previously studied for the observed data in Figure~\ref{fig:footprint_stats}. The comparison of observed and simulated winds is very good; in particular it replicates the feature that the largest magnitudes tend to be found in the region southwest of the storm centre. The local minimum that occurs near the storm centre as a result of small pressure gradients is also captured. The upper quantiles of the spatial distribution are slightly more dispersed compared to the observed characteristics; however, we are satisfied that the large-scale features have been captured by the model. Figure~\ref{fig:spatial_sim} also shows that high magnitude events can be generated by the model on the outer edges of the domain. These are rare occurrences, but we believe these to be attributed to the detection of spurious events in the feature extraction algorithm. Section~\ref{sec:discussion} discusses possible improvements to the algorithm so that the detection of spurious events is minimised. \\

\begin{figure}
\centering
\includegraphics[width=12cm]{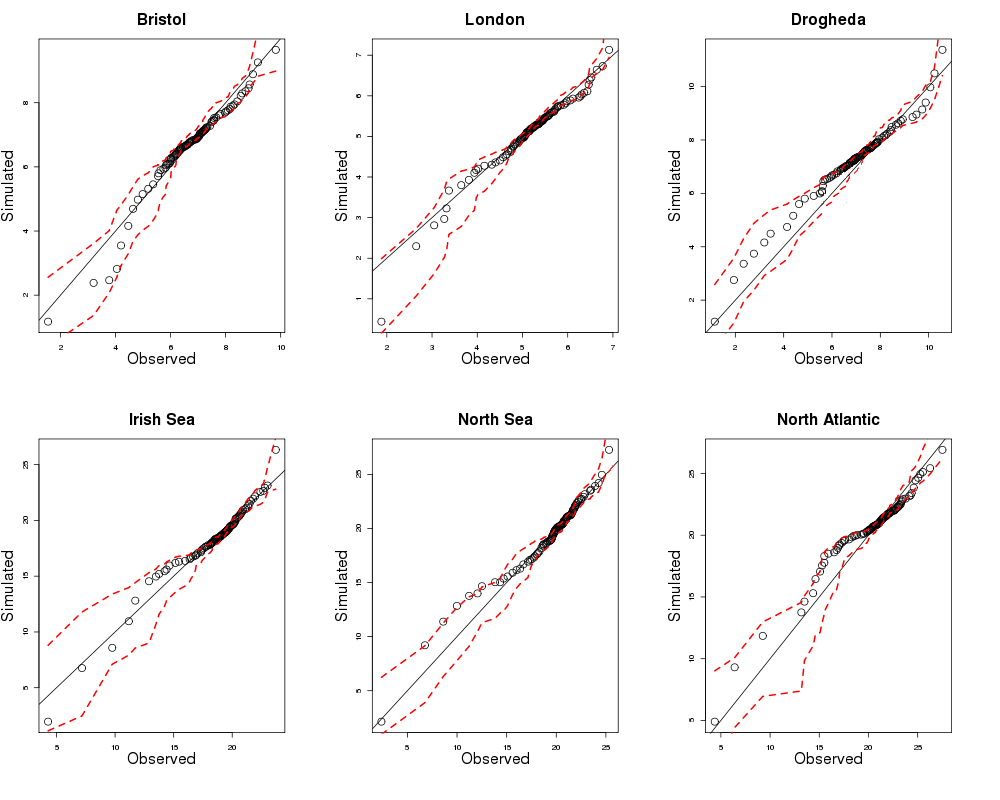}
\caption{QQ plots, with $95\%$ tolerance intervals, at six locations comparing the distribution of wind speeds from the observed and simulated windstorms.}
\label{fig:qq_winds}
\end{figure}

By simulating windstorms relative to synthetic tracks, our model also allows us to perform an Eulerian analysis at different locations over the North Atlantic and Europe. We assess how our model captures the distribution of extreme winds at a number of locations to examine whether our approach succeeds in generating physically realistic synthetic values at these different sites. We compare our simulated values with winds that are contained within the observed footprints at each site. QQ plots for six locations, three on land and three at sea, are shown in Figure~\ref{fig:qq_winds}. The $100$-year return level (not shown), estimated from our model for the spatial regions discussed in Section~\ref{subsec:joint_risk}, compares favourably with estimates from the marginal model in~(\ref{eq:marginal_model}), with an average percentage error of $2.4\%$ over all cells. This result demonstrates that our model can be used to assess marginal risk at different locations for events beyond the range of the observational record. 
%
%
%

\subsection{Joint risk from windstorms}
\label{subsec:joint_risk}
We use our approach to estimate quantities related to joint risk, that is, the probability that multiple locations are affected by extreme wind speeds simultaneously. As our model captures the spatial extent of these meteorological events, we should therefore capture the risk of multiple locations experiencing extreme wind speeds from the same storm. The results are based on 50,000 extratropical cylcone tracks generated using the track model of \citet{sharkey2017tracking}.  For each track, a windstorm is simulated using the windstorm model of Section \ref{sec:methodology}. This dataset represents approximately $600$ years of data, under the assumption that there are on average 811 windstorms per year as found in the observed record. \\ 

\begin{figure}[h!]
\centering
\includegraphics[width=8cm]{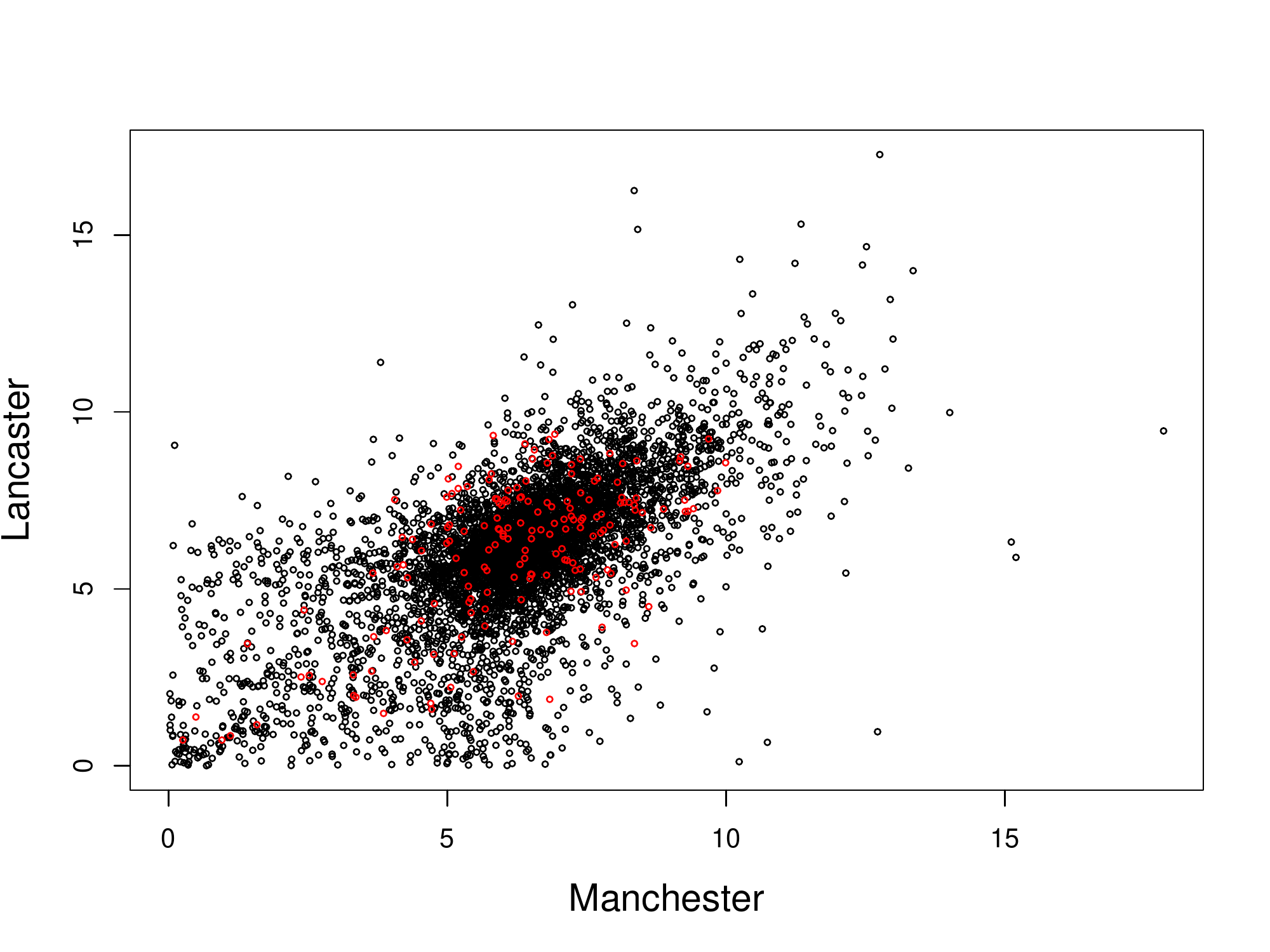}
\includegraphics[width=8cm]{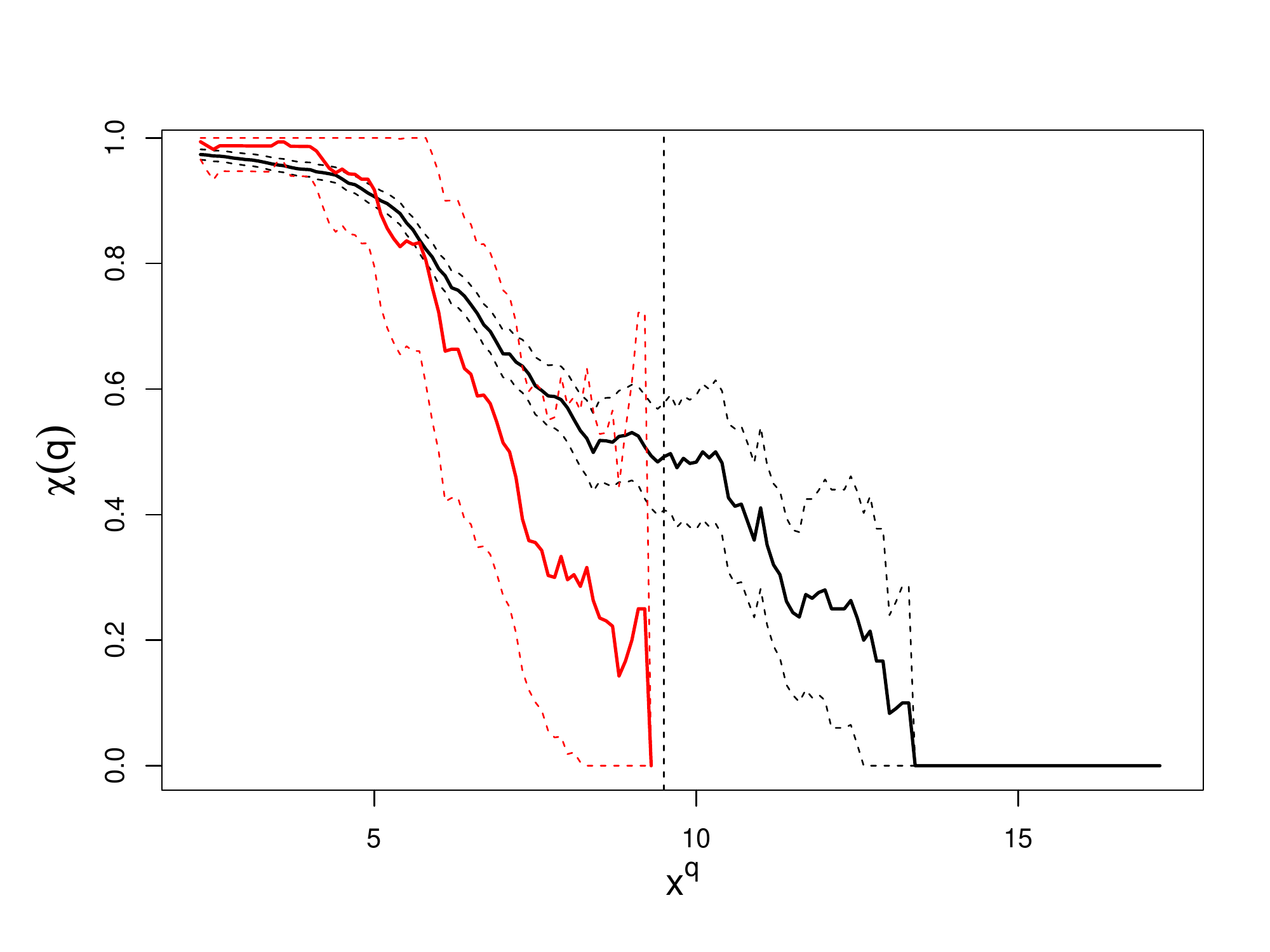}
\caption{Scatter plot (left) showing observed (red) and simulated (black) wind speeds on an Exp$(1)$ scale in Manchester and Lancaster. Estimates of $\chi(q;s_1,s_2)$ measuring extremal dependence between these locations as a function of $x^q$ (right) using the observed (red) and simulated (black) data, with $95\%$ binomial confidence intervals using an effective sample size. The vertical line denotes the maximum observed wind speed on the Exp$(1)$ scale.} 
\label{fig:chi}
\end{figure}

Figure~\ref{fig:chi} (left panel) compares the joint behaviour on common Exp$(1)$ margins between wind speeds in Lancaster and Manchester, two cities in northern England 73km apart. The joint correlation structure is largely captured by the model, which has the added benefit of being able to simulate joint events of magnitudes beyond the range of the observation record. One way of summarising the joint extremal behaviour of a process at arbitrary locations $s_1$ and $s_2$ is to estimate the quantity $\chi(q; s_1,s_2)$ \citep{coles1999dependence}, defined as
\[ \chi(q; s_1,s_2) = \Pr(X^E (s_2,t) > x^q \mid X^E (s_1,t) > x^q), \]
where $x^{q}$ is the $100q\%$ quantile of the common Exp$(1)$ distribution. Estimates of the quantity $\chi(q; s_1, s_2)$ obtained from the observed and simulated data are shown for a range of $x^{q}$ in Figure~\ref{fig:chi} (right panel), where $s_1$ and $s_2$ are chosen to be sites in Lancaster and Manchester respectively. Estimates from the data and the large simulated sample from the model are obtained as conditional proportions. Figure~\ref{fig:chi} shows that extremal dependence tends to decrease as the magnitude of the event increases. Here, $95\%$ binomial confidence intervals are used to assess the uncertainty for the observed and the Monte Carlo uncertainty for the simulated data. To obtain these intervals, we use an effective sample size $n\theta(x^q)$ defined in terms of the sample size $n$ and a threshold-based extremal index $\theta(x^q)$ \citep{ferro2003inference,eastoe2012modelling} to account for temporal dependence. For the model-based estimates, the confidence intervals do not represent the uncertainty due to the model parameter estimation. A fuller assessment of model uncertainty can be obtained using a parametric bootstrap, which would have the effect of widening the model-based confidence intervals. Despite not representing the full uncertainty in the model-based estimates, it is clear from the overlapping of the confidence intervals that there is little difference between data and model-based estimates of $\chi(q; s_1,s_2)$ here over $q$ within the range of the observed data. Critically though, this figure also illustrates how our model allows estimation of $\chi(q; s_1,s_2)$ beyond the range of the observational record, indicating that $\chi(q;s_1,s_2)$ continues to decay to $0$ beyond the observed data. \\

\begin{figure}[h!]
\centering
\includegraphics[width=8cm]{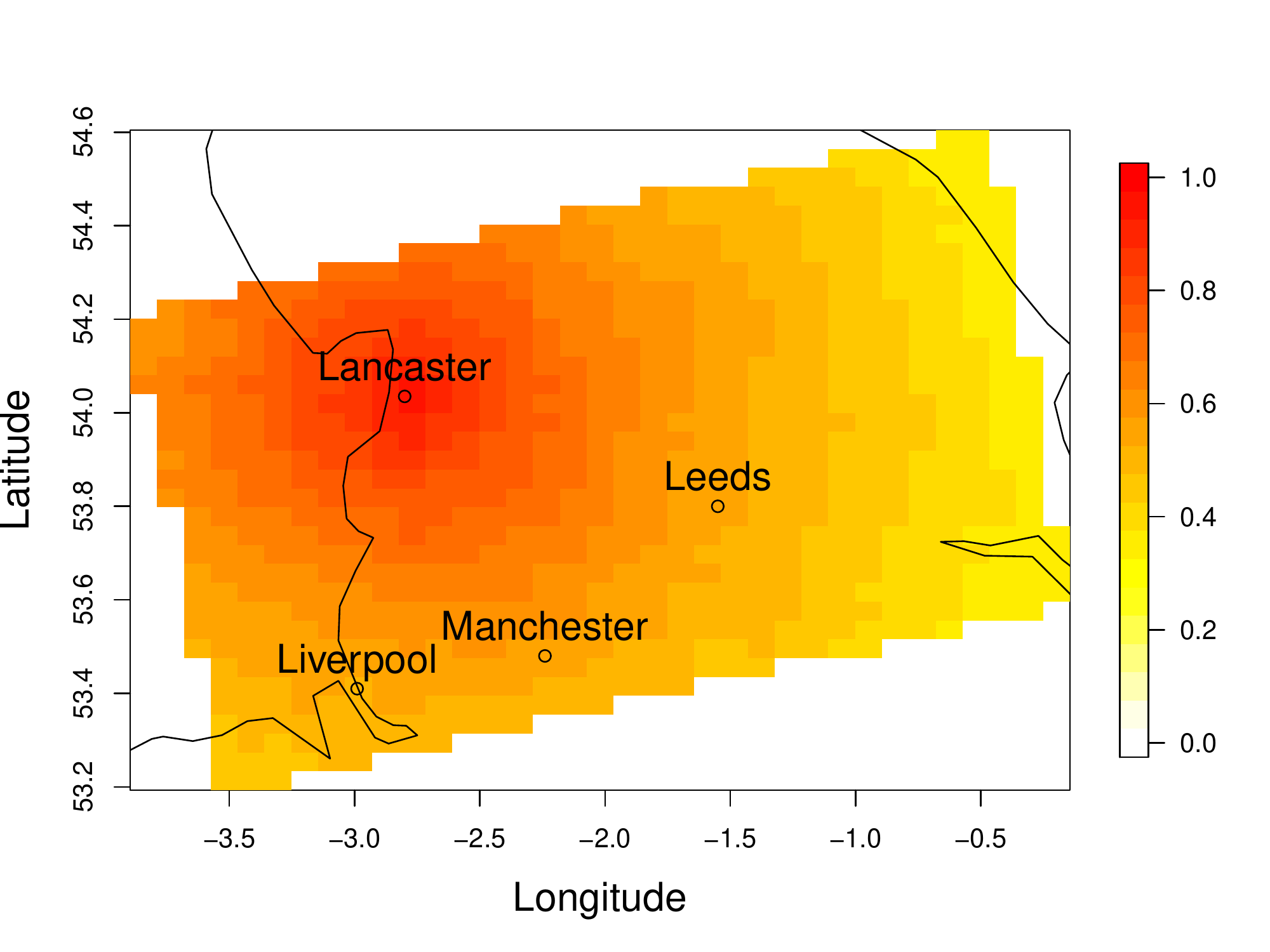}
\includegraphics[width=8cm]{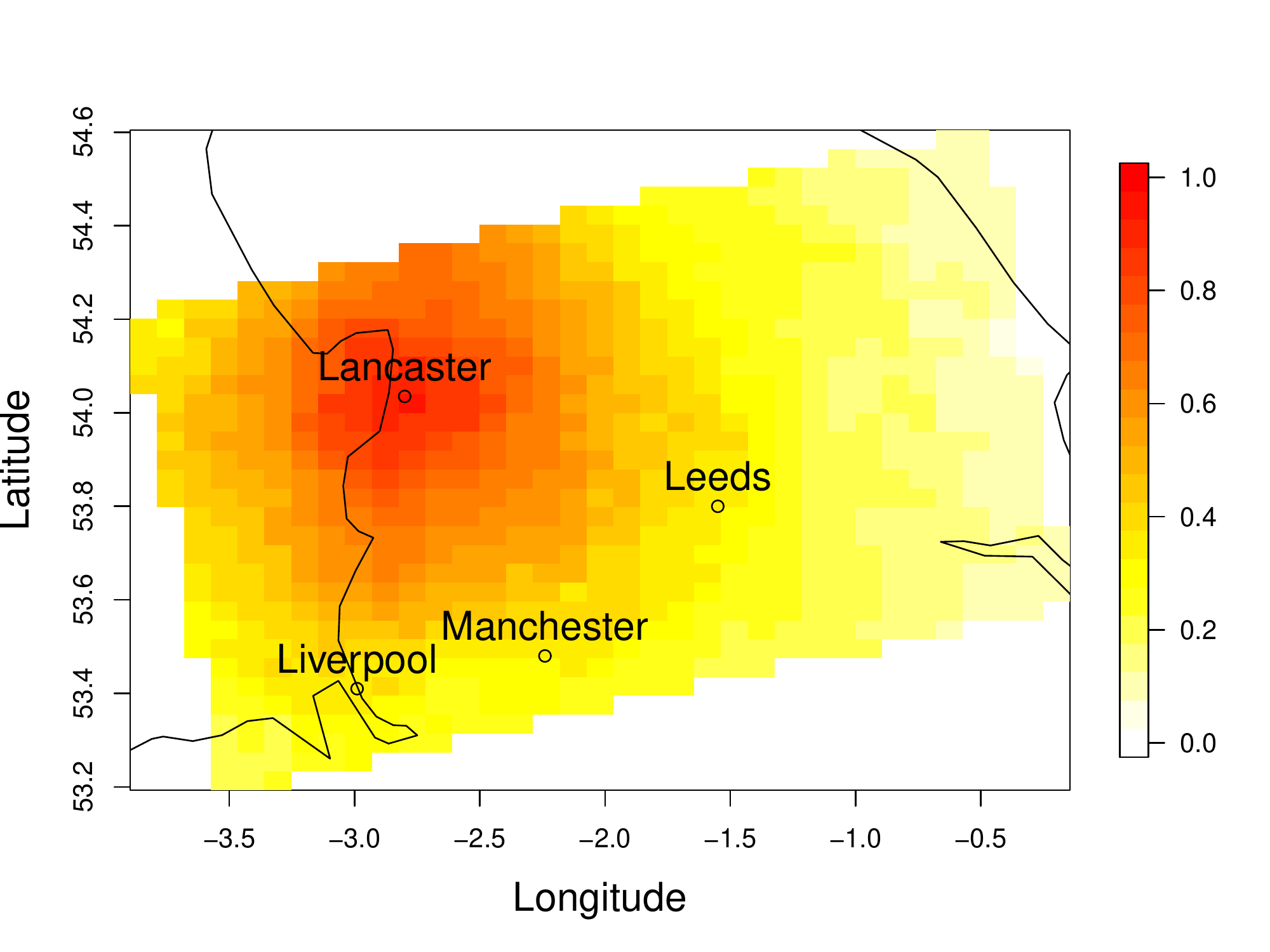}
\includegraphics[width=8cm]{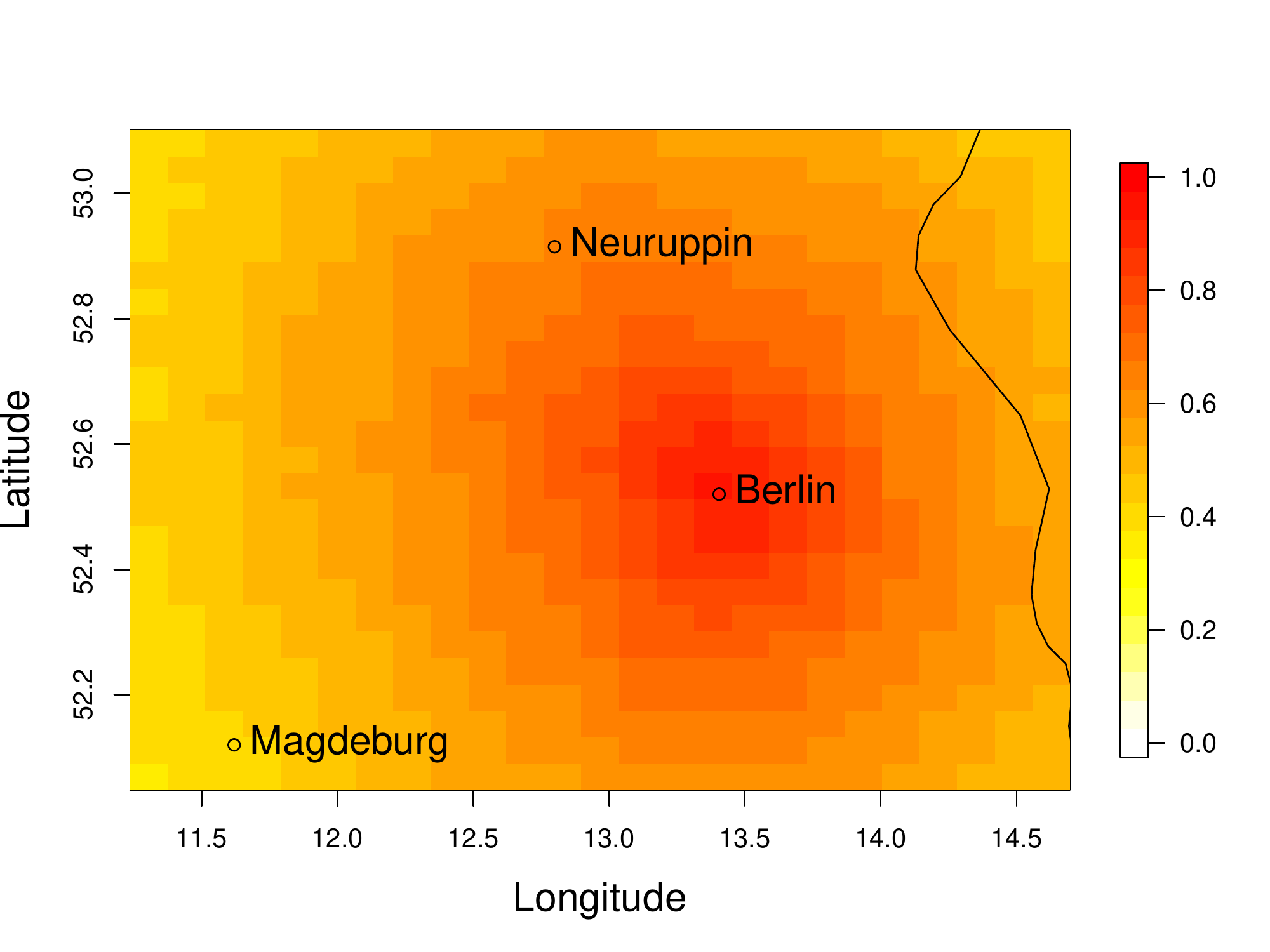}
\includegraphics[width=8cm]{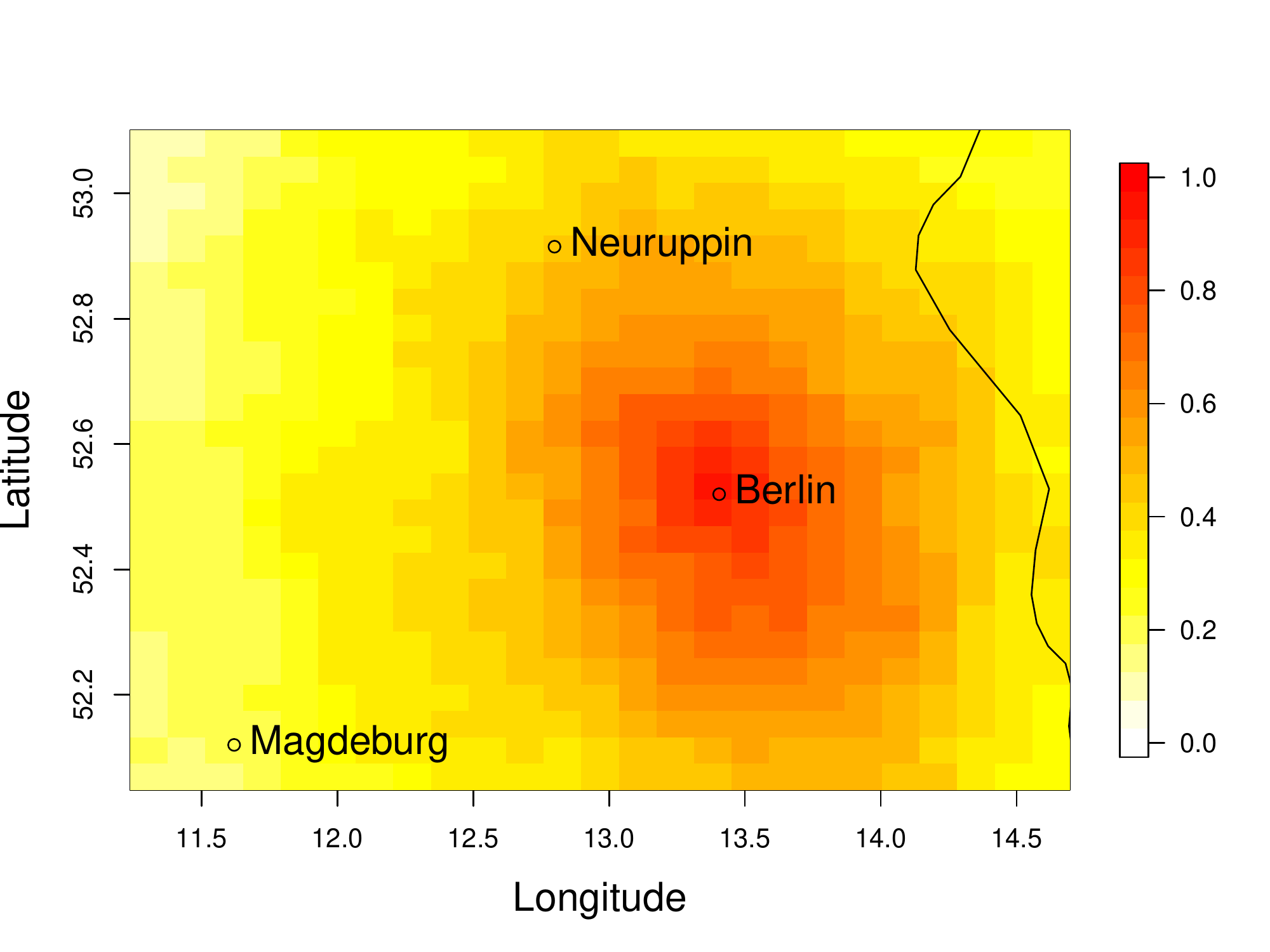}
\caption{Estimates of $\chi(q;s_1,s_2)$ for northern England (top) and eastern Germany (bottom) conditioning on a critical value $x^q$, where $s_1$ is the cell where Lancaster (top) and Berlin (bottom) are located. In the left panels, $x^q$ is taken to be the $90\%$ quantile, while $x^q$ is taken to be the $10$-year return level in the right panels. Both regions are of equal size. }
\label{fig:chi_space}
\end{figure}

Estimating $\chi(q;s_1,s_2)$ at a fixed critical level $x^q$ at a set of sites $s_2$ can allow us to explore the spatial extent of extreme events. Figure~\ref{fig:chi_space} (top panels) shows $\chi(q; s_1,s_2)$ calculated across a number of locations in northern England, with $s_1$ being Lancaster in this instance. We explore two cases where $x^q$ is chosen to the $90\%$ quantile and the $10$-year return level at each site. In particular, we see that the probability surface decays more steeply as $|s_2-s_1|$ increases for the more extreme events. We also see that Liverpool, Manchester and Leeds are more likely than not to experience an event on the $90\%$ quantile simultaneously to Lancaster; however, this scenario is less likely for an event corresponding to the $10$-year return level. Similarly, in Germany, as shown in the bottom panels in Figure~\ref{fig:chi_space}, the probability of experiencing an extreme windstorm event simultaneously with Berlin decreases as the event becomes more extreme. The spatial extremal dependence is slightly stronger for Berlin than Lancaster, as might be expected given Berlin is more inland on a large land mass. In both regions, there is little evidence of anisotropy in the extremal dependence structure, with perhaps some indication of stronger dependence in the northwest-southwest direction centred at Berlin. The results in Figures~\ref{fig:chi} and~\ref{fig:chi_space} illustrate how the spatial extent of an extreme windstorm event becomes more localised as the magnitude increases. This result implies that in the limit, extreme values at each location tend not to occur simultaneously, which corresponds to the property of asymptotic independence. Models for asymptotic dependence, that is, when $\chi(q;s_1,s_2) \rightarrow c > 0$ as $q \rightarrow 1$ for $s_1 \neq s_2$, lead to extreme values tending to occur simultaneously, are well-established but tend to over-estimate the probability of extreme joint events given that the underlying process is asymptotically independent, i.e., when $c=0$. Models that capture asymptotic independence are less well-established; see \citet{ledford1996statistics}, \citet{heffernan2004conditional}, \citet{wadsworth2012dependence} and \citet{winter2016modelling} for some examples. We have shown that our model captures the property of asymptotic independence over space, while accounting for the complex non-stationarity of the extratropical cyclone system.

\section{Discussion}
\label{sec:discussion}


We have presented a novel approach to modelling windstorms in a Lagrangian framework. We described two models; first, for the evolution and development of the footprint relative to the storm track, and second, for the spatial distribution of extreme winds within the footprint. The Lagrangian framework allows us to pool information regarding events over the spatial domain being studied, which allows extrapolation of the characteristics of windstorms over space. The model provides a mechanism for generating synthetic windstorm events, the analysis of which allows improved estimation of joint risk associated with extreme windstorms over Europe. \\

There are, however, opportunities to improve the performance of the model. While the feature extraction approach introduced in Section~\ref{subsec:feature} appears to extract the main features of a windstorm, steps could be taken to improve the robustness of this procedure. Firstly, one could conduct a sensitivity study on the choice of threshold $v$, which controls the level under which the wind speed fields are masked. Ideally, one should choose a high enough value of $v$ such that convective events and non-extreme winds are masked, but small enough so that the localised features of the windstorm are not excluded. ``Sting jets", meteorological phenomena associated with rapidly developing storms, can produce damaging winds on very small spatial scales \citep{baker2009sting,hewson2015cyclones} and therefore it is important that the extraction algorithm should not exclude such features. \\

The feature extraction algorithm could also be improved to minimise the detection of spurious footprints that may not be generated by the extratropical cyclone. We see examples of this through high magnitude events occurring on the outer edges of the domains in Figure~\ref{fig:footprint_stats}. After the spatio-temporal filtering step, our algorithm selects the largest cluster in size to define the footprint. One could alternatively define a score function that incorporates multiple criteria in a detection strategy, which could be motivated by physical intuition regarding the structure of a windstorm. For example, the score function could give a higher weight to clusters with a bearing relative to the storm centre closest to southwest, where most footprints seem to occur. To induce some smoothness between consecutive time steps, one could assign a higher weight to a cluster such that the Euclidean distance between the position of this cluster and the selected cluster at the previous time step is minimised. Exploration of different score functions could add valuable improvements to our feature extraction algorithm, and ultimately, the model. \\  

Our conditional kernel strategy for modelling $\{\boldsymbol{Z}_t\}$ appears to perform well. However, we could alternatively model the extremal behaviour of $W_t$ using the approaches described in \citet{winter2017kth} and \citet{sharkey2017tracking}, whereby a GPD model is defined above a high threshold and the extremal temporal dependence structure is modelled using a $k$th order extremal Markov process. This approach stems from the conditional multivariate extreme value methodology of \citet{heffernan2004conditional}. We did not implement this approach as part of this study due to the additional complexity involved and that extrapolation occurred naturally through the features of our model. In addition, the observed values of $W_t$ correspond to upper tail values of $X_i$, the distribution of wind speeds in cell $i$. Because we have a large dataset of observations in the upper tail, we felt that standard statistical modelling approaches were sufficient in this case. However, investigation of the benefits of imposing an extremal temporal dependence structure on the upper tail of $W_t$ represents an interesting avenue of future research. \\

The potential future risk associated with extreme windstorms is of pressing concern in addition to how their characteristics might be affected by a changing climate. For the North Atlantic, previous studies have indicated the winter storm track will potentially increase over the UK and northern Europe but eliciting this signal is difficult \citep{zappa13future}.  This uncertainty, in combination with the low probability of a cyclone producing extreme winds makes estimating future windstorm risk very challenging. The windstorm model presented here and the track model of \citet{sharkey2017tracking} together provide a new tool that can be applied to future climate simulations to potentially provide improved estimates of such risk.

\section*{Acknowledgements}
The authors are thankful for the contributions of two referees that vastly improved the quality of the paper. The authors gratefully acknowledge the support of the EPSRC funded EP/H023151/1
STOR-i Centre for Doctoral Training, the Met Office and EDF Energy. We extend our thanks to Hugo Winter for helpful discussions and support. We thank Kevin Hodges for the storm track data. Simon Brown was supported by and Paul Sharkey partially supported by the Joint UK BEIS/Defra Met Office Hadley Centre Climate Programme (GA01101).
\appendix
\section*{Appendix}
\section{Conditional kernel density estimation}
\label{App:sim_kernel}
Consider an arbitrary $d$-dimensional random vector $\boldsymbol{Z} = (Z_1, Z_2, \hdots, Z_d)$, which is observed $n$ times $\boldsymbol{z}^{(1)}, \boldsymbol{z}^{(2)}, \hdots, \boldsymbol{z}^{(n)}$. As a way of estimating $f(\boldsymbol{z})$, the joint probability density of $\boldsymbol{Z}$, we define the multivariate kernel density estimator as 
\begin{equation}
 \hat{f}(\boldsymbol{z}) = \frac{1}{n} \sum_{i=1}^{n} K_{H} \left(\boldsymbol{z} - \boldsymbol{z}^{(i)} \right),
 \label{eq:mkde} 
\end{equation}
where $K_H$ is the kernel function and $H$ denotes the bandwidth matrix which is symmetric and positive-definite. For our purposes, we choose $K_H$ to be the multivariate Gaussian density function
\begin{equation}
 K_{H} (\boldsymbol{z}) = {(2 \pi)}^{-d/2} {|H|}^{-1/2} \exp\left\{-\frac{1}{2} \boldsymbol{z}^{T} H^{-1} \boldsymbol{z} \right\}
 \label{eq:kernel}
 \end{equation}
and the bandwidth matrix $H$ chosen to be proportional to the rule-of-thumb selection of \citet{scott2015multivariate}. The bandwidth matrix $H$ can be chosen to be diagonal or oriented. \\

Let $\boldsymbol{Z}$ be decomposed such that $\boldsymbol{Z} = (\boldsymbol{Z}_{-m},\boldsymbol{Z}_m)$. Consider the case when values $\boldsymbol{Z}_{-m} = \boldsymbol{z}_{-m}$ have been observed and we wish to estimate the conditional density of $\boldsymbol{Z}_{m}$ given these values. We can then define the conditional kernel density estimator as
\begin{equation} \hat{f}(\boldsymbol{z}_m | \boldsymbol{z}_{-m} ) = \sum_{i=1}^n w_i (\boldsymbol{z}_{-m}) K_{H} \left(\boldsymbol{z}_m - \boldsymbol{z}_m^{(i)} \middle| \boldsymbol{z}_{-m} - \boldsymbol{z}_{-m}^{(i)} \right),
\label{eq:ckde}
\end{equation}
where 
\[
w_i (\boldsymbol{z}_{-m}) = \frac{K_{H} \left( \boldsymbol{z}_{-m} - \boldsymbol{z}_{-m}^{(i)} \right)}{ \sum_{j=1}^{n} K_{H} \left( \boldsymbol{z}_{-m} - \boldsymbol{z}_{-m}^{(j)} \right)},
\]
where $K_{H} ( \cdot )$ is the multivariate Gaussian kernel function and $K_{H} ( \cdot \mid \cdot )$ is the conditional Gaussian kernel function with bandwidth matrix $H$ as defined in equation~(\ref{eq:kernel}). Let $H$ be partitioned such that 
\[ H = \begin{bmatrix}
H_{m,m} & H_{m,-m} \\ H_{-m,m} & H_{-m,-m}
\end{bmatrix}. \]

Conditional on having observed $\boldsymbol{z}_{-m}$, we choose a tuple $\boldsymbol{z}^{(i)}$ with probability $w_i (\boldsymbol{z}_{-m})$. Then we simulate 
\begin{equation}
 \boldsymbol{Z}_m | (\boldsymbol{Z}_{-m} = \boldsymbol{z}_{-m}) \sim \mathcal{N}(\bar{\boldsymbol{\mu}}, \bar{\boldsymbol{\Sigma}}),
 \label{eq:kersim}
\end{equation}
where $\bar{\boldsymbol{\mu}} = z_m^{(i)} + H_{m,-m} H_{-m,-m}^{-1} (\boldsymbol{z}_{-m} - \boldsymbol{z}_{-m}^{(i)}$) and $\bar{\boldsymbol{\Sigma}} = H_{m,m} - H_{m,-m}H_{-m,-m}^{-1} H_{-m,m}$. \\

\bibliographystyle{apalike}  
\bibliography{wind_speed_paper}
\end{document}